\pdfoutput=1
\documentclass[10pt]{article}
\usepackage[utf8]{inputenc}
\usepackage{cite}
\usepackage{hyperref}
\hypersetup{colorlinks=true,linkcolor=blue2,citecolor=red2,urlcolor=green2,pdfencoding=auto,linktocpage}
\usepackage{amsmath,amssymb,amsbsy,amstext,amsthm,simplewick,amsfonts}
\usepackage{mathrsfs}
\usepackage{graphicx}
\usepackage{wrapfig}
\usepackage{upgreek}
\usepackage{bm} 
\usepackage{framed}
\usepackage{bbm}
\usepackage{textcomp}
\usepackage{adjustbox}
\usepackage{makecell}
\usepackage{tcolorbox}
\usepackage{empheq}
\usepackage[normalem]{ulem}
\usepackage{enumitem}
\usepackage{braket} 
\usepackage{array}
\usepackage{dsfont}
\usepackage{physics}
\usepackage{ulem}
\usepackage{xcolor}
\usepackage{caption}
\usepackage{subcaption}
\usepackage{tocloft}
\usepackage{tikz}

\usepackage[framemethod=default]{mdframed}

\newmdenv[backgroundcolor=gray!15,%
skipabove=5pt,%
skipbelow=5pt,%
leftmargin=2pt,%
rightmargin=2pt,%
innertopmargin=-6pt,%
innerbottommargin=5pt,%
innerleftmargin=5pt,%
innerrightmargin=5pt,%
splittopskip=0pt,%
splitbottomskip=0pt,%
linewidth=0pt,%
nobreak=true]%
{keyeqn}

\graphicspath{{Fig/}}

\definecolor{red2}{RGB}{214, 39, 40}
\definecolor{green2}{RGB}{0,170,0}
\definecolor{blue2}{RGB}{0,100,200}
\definecolor{magenta2}{RGB}{191,64,191}
\definecolor{purple2}{RGB}{112,48,160}
\definecolor{orange2}{RGB}{255,192,0}

\def\fnl{f_{\rm NL}}

\def\d{\mathrm{d}}

\def\Re{\mathrm{Re}\,}

\newcommand{\Planck}{\textit{Planck}}
\newcommand{\fnlA}{f_{\rm NL}^{(\alpha)}}
\newcommand{\mri}{\mathrm{i}}
\newcommand{\mrf}{\mathrm{f}}

\newcommand{\ex}[1]{\left\langle #1 \right\rangle}

\tikzset{dS diagram/.style={execute at begin picture={%
         \draw (-0.5*  \pgfkeysvalueof{/tikz/dS/width},0) -- (0.5 * \pgfkeysvalueof{/tikz/dS/width}, 0);
         \node (X) at (0,-\pgfkeysvalueof{/tikz/dS/aspect}*\pgfkeysvalueof{/tikz/dS/width}) {\phantom{X}};
},baseline={(X.base)}},vertex/.style={circle,fill,inner sep=1.5pt,node contents={}},
dS/.cd,width/.initial=3cm,aspect/.initial=0.3}
\newenvironment{dSdiagram}[1][]{\begin{tikzpicture}[dS diagram,dS/.cd,#1]}{\end{tikzpicture}}
\newcommand{\dSwidth}{\pgfkeysvalueof{/tikz/dS/width}}

\numberwithin{equation}{section}

\allowdisplaybreaks[1]
\begin{document}

\begin{titlepage}
	\setcounter{page}{1} \baselineskip=15.5pt 
	\thispagestyle{empty}


     \begin{center}
		{\fontsize{18}{18}\centering {\bf{A Cosmological Tachyon Collider:\\[0.2cm] Enhancing the Long-Short Scale Coupling}}}
	\end{center}

 
	\vskip 18pt
	\begin{center}
		\noindent
		{\fontsize{12}{18}\selectfont Ciaran McCulloch\footnote{\tt cam235@cam.ac.uk}$^{,a}$, Enrico Pajer\footnote{\tt enrico.pajer@gmail.com}$^{,a}$, and Xi Tong  \footnote{\tt xt246@cam.ac.uk}$^{,a}$}
	\end{center}
	
	\begin{center}
		\vskip 12pt
		$^{a}$ \textit{Department of Applied Mathematics and Theoretical Physics, University of Cambridge,\\Wilberforce Road, Cambridge, CB3 0WA, UK} \\
	\end{center}
	
	\vskip 50pt
	\noindent\rule{\textwidth}{0.4pt}
	\noindent \textbf{Abstract} ~~ The squeezed limit of the primordial curvature bispectrum is an extremely sensitive probe of new physics and encodes information about additional fields active during inflation such as their masses and spins. In the conventional setup, additional fields are stable with a positive mass squared, and hence induce a decreasing signal in the squeezed limit, making a detection challenging.
	
    Here we consider a scalar field that is temporarily unstable by virtue of a transient tachyonic mass, and we construct models in which it is embedded consistently within inflation. Assuming IR-finite couplings between the tachyon and the inflaton, we find an exchange bispectrum with an enhanced long-short scale coupling that grows in the squeezed limit parametrically faster than local non-Gaussianity. Our approximately scale-invariant signal can be thought of as a \textit{cosmological tachyon collider}. 
    
    In a sizeable region of parameter space, the leading constraint on our signal comes from the cross correlation of $\mu$-type spectral distortions and temperature anisotropies of the microwave background, whereas temperature and polarization bispectra are less sensitive probes. By including anisotropic spectral distortions in the analysis, future experiments such as CMB-S4 will further reduce the allowed parameter space. 
    

	\noindent\rule{\textwidth}{0.4pt}
	
	
\end{titlepage} 


\newpage
\setcounter{page}{2}
{
	\tableofcontents
}




\section{Introduction}\label{IntroSect}

\setcounter{footnote}{0}

The study of the primordial universe is also a study of physics at exceedingly higher energy scales, by virtue of the cosmic expansion. It is widely believed that there existed a stage of cosmic inflation during which the universe expanded enormously, with a background geometry well described by a quasi-de Sitter spacetime \cite{Starobinsky:1980te,Guth:1980zm,Linde:1981mu,Albrecht:1982wi}. The fast expansion of spacetime serves as a natural accelerator for massive fields during inflation, and is capable of producing particles with masses up to the Hubble scale during inflation, which can be as high as $H \leq 5\times 10^{13}$ GeV. These massive particles can interact and decay into massless inflatons and gravitons, sourcing characteristic signals in the primordial non-Gaussianities of curvature and tensor perturbations in the Cosmic Microwave Background (CMB) and Large Scale Structure (LSS). In recent years, this cosmological collider paradigm \cite{Chen:2009zp,Baumann:2011nk,Noumi:2012vr,Arkani-Hamed:2015bza,Lee:2016vti} has received much attention and witnessed fast developments across aspects of its 
theory \cite{Baumann:2022jpr}, 
phenomenology \cite{Chen:2016uwp,Chen:2018xck,Chen:2023txq,An:2017hlx,Hook:2019zxa,Kumar:2019ebj,Bodas:2020yho
,Alexander:2019vtb,Liu:2019fag,Tong:2022cdz,Wang:2019gbi,Cui:2021iie,Reece:2022soh,Aoki:2020zbj,Pinol:2021aun,Jazayeri:2023xcj}, and observational prospects \cite{Meerburg:2016zdz,MoradinezhadDizgah:2018ssw,Kogai:2020vzz,Kalaja:2020mkq,Cabass:2022oap,Floss:2022grj}.\\

\begin{figure}[h!]
   \centering
   \includegraphics[width=0.6\textwidth]{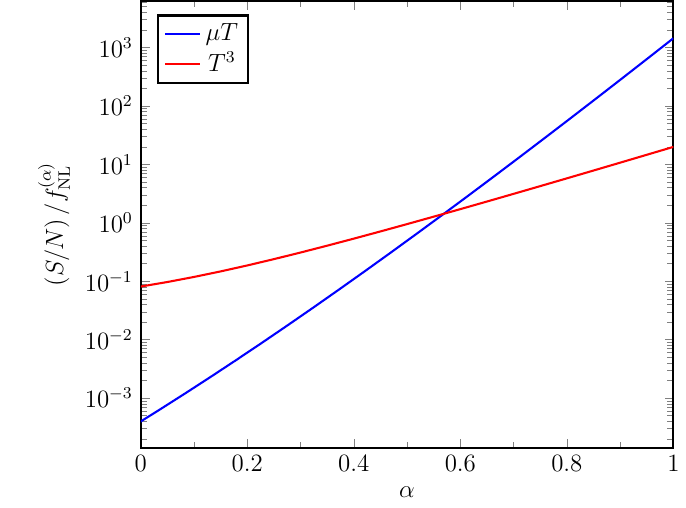}
   \caption{Signal-to-noise of the amplitude $\fnlA$ as a function of the parameter $\alpha$ (defined in \eqref{eq:intro squeezed bispectrum}). The $\mu T$ line assumes a PIXIE-like experiment \cite{Kogut:2011xw,Pajer:2012vz}, while $T^3$ refers to \Planck's sensitivity to the CMB-scale bispectrum. When $\alpha$ is not close to zero, the slope of the $\mu T$ line is roughly $ \log_{10} (k_\mu / k_\text{CMB}) \sim 6.7$ where $k_\text{CMB}$ is the longest scale contributing to the CMB temperature constraint and $k_\mu$ (not to be confused with a four-vector) the characteristic scale contributing to $\mu$ distortion. The slope of the $T^3$ line is 
   controlled by the range of scales visible in the CMB anisotropies, roughly $ \log_{10} (l_\text{max}/l_\text{min}) \sim 3$.} 
   \label{SNIntroPlot}
\end{figure}

More specifically, the relevant observables are boundary $n$-point correlators of massless curvature and tensor perturbations evaluated at the end of inflation. The defining feature of cosmological collider signals is a non-analytic scaling behaviour in the kinematic region in which one of the momenta of a bispectrum is much smaller than the other two, the so-called squeezed limit. In particular, the squeezed limit of the curvature bispectrum scales as $(k_L/k_S)^{3/2-\nu}$ for $k_L\ll k_S $, where the labels $S$ and $L$ stand for short and long modes, respectively. For light fields, namely fields with mass $m\leq (3/2)H$, one finds that $0\leq \nu\leq 3/2$ is a real number and the bispectrum attenuates as a power law fashion in the squeezed limit \cite{Chen:2009zp}. For heavy fields, $\nu=i|\nu|$ is purely imaginary and the bispectrum oscillates as the logarithm of the momentum ratio in the squeezed limit \cite{Noumi:2012vr}. In both cases, the exponent $\nu$ is determined by the mass of the extra field that mediates the interaction. Therefore, measurements of the power-law exponent or the oscillation frequency in principle probe the mass spectrum during inflation.\\

The decay in the squeezed limit engenders a series of challenges. On the theory side, the signals from the exchange of stable massive particles living in the unitary irreducible representations of the de Sitter group always decay in the squeezed limit (e.g. $\Re(3/2-\nu)\geq 0$ in the bispectrum case), which suppresses the overall signal-to-noise ratio. For heavy fields, there is an additional Boltzmann factor $e^{-\pi m/H}$ that cuts off the signal for masses well above the Hubble scale \cite{Arkani-Hamed:2015bza}. On the observation side, the detection of cosmological collider signals requires observing modes of very different wavelength. In the CMB temperature and polarisation maps, short scales are erased by diffusion damping and by the thickness of the last scattering surface. Moreover, large scale structure surveys are also limited to large scales by the strongly non-perturbative dynamics induced by gravitational collapse on short scales. It is undoubtedly a challenge to reach meaningful theoretical benchmarks given current observational capabilities \cite{Kalaja:2020mkq}.\\

On the other hand, the decay of correlations in the squeezed limit can be understood more generally as the decoupling of modes in the presence of a hierarchy of scales. This notion of UV-IR decoupling, despite being natural, is  more subtle in an expanding universe. It is interesting to examine if there could be some non-trivial UV-IR mixing that gives rise to \textit{growing} correlation between physics separated by a greater hierarchy of scales. If there is indeed a mechanism capable of producing correlation growing with the scale hierarchy, the remaining question is to find the natural observational probe to verify such a mechanism.\\

Motivated by the challenges and questions above, in this work, we propose a cosmological tachyon collider model that partially addresses them. By introducing a scalar field $\chi$ with temporarily negative squared mass, which weakly interacts with the inflaton field, we show that the tachyonic dynamics produce correlations of curvature perturbations that grow with the scale hierarchy. In the curvature bispectrum, this results in a signal that can generically be parametrised as
\begin{keyeqn}
    \begin{align}
	\ex{\zeta(k_S)\zeta(k_S)\zeta(k_L)}'=(2\pi^2\Delta_\zeta^2)^2 \,f_{\rm NL}^{(\alpha)}\, \frac{1}{k_S^3 k_L^3}\left(\frac{k_S}{k_L}\right)^\alpha~,\quad \alpha>0~,
    \label{eq:intro squeezed bispectrum}
    \end{align}
\end{keyeqn}
in the squeezed limit $k_S/k_L\gg 1$, where $f_{\rm NL}^{(\alpha)}$ described the amplitude of the signal and the scaling exponent $\alpha$ is related to the tachyon mass by $\alpha=\tilde{\nu}-3/2$ and $\tilde{\nu}^2=9/4+\tilde{m}^2/H^2$. Here we have stripped a momentum-conserving delta function and $\Delta_\zeta^2\simeq 2\times 10^{-9}$ is the primordial curvature power spectrum\footnote{\label{foot}The parameter $\Delta_\zeta^2$ is equivalent to the parameter $A_s$ of \cite{Planck:2018vyg}, although we assume exact scale invariance for simplicity, so the primordial power spectrum is of the form $\ev{\zeta(\vb k) \zeta(-\vb k)}' = 2\pi^2\Delta_\zeta^2/k^3$. The normalisation of the bispectrum is as in \cite{Planck:2019kim}, except that that work used the bispectrum of the Bardeen potential $\Phi \approx 3\zeta/5$. See \cite{Sohn:2023fte} for a description of the various normalisation conventions for the primordial power spectrum. }. As the momentum ratio $k_S/k_L$ increases, the signal also increases as a power law. In particular, the shape of non-Gaussianity peaks in the squeezed limit \textit{faster} than local non-Gaussianity, which corresponds to $\alpha=0$. This growth is eventually cut off by the restoration of a positive squared mass for the $\chi$ field. We discuss a few different model realisations to end the transient tachyonic phase and examine their theoretical constraints. Our setup generally allows for $0<\alpha\lesssim 1$ and a scale hierarchy of $k_S/k_L\lesssim 10^7$. 

We also propose to detect the above signal in the two-point cross correlation of anisotropies in the CMB temperature fluctuations $T$ and $\mu$-type spectral distortion \cite{Pajer:2012vz}. We analyse the signal-to-noise ratio for $\mu T$ and compare it to the CMB ($T,E$)-bispectrum in Figure \ref{SNIntroPlot}. One can see that the signal-to-noise ratio for $\mu T$ grows exponentially with $\alpha$ and exceeds that in the CMB bispectrum for $\alpha\gtrsim 0.5$, making it the most sensitive probe of highly ultra-squeezed non-Gaussianity. This happens because the signal-to-noise in the $\mu T$ measurement grows roughly as $(k_\mu / k_\text{CMB})^\alpha \sim 10^{6.7 \,\alpha}$, where $k_\mu$ is the characteristic momentum scale probed by the $\mu$ distortions and $k_{\text{CMB}}$ is the longest scale contributing to the CMB temperature anisotropy constraint. Here, we take $k_\mu \approx 740 \, \mathrm{Mpc}^{-1}$, following \cite{CMB-S4:2023zem}, and $k_\text{CMB} \approx 2 /( 14 \mathrm{Gpc})$, corresponding to the CMB quadrupole. On the other hand, the signal-to-noise in the CMB ($T,E$)-bispectrum measurement grows roughly as $(l_\text{max}/l_\text{min})^\alpha \sim 10^{3\,\alpha}$, where $l_\text{max}$ and $l_\text{min}$ correspond to the shortest and longest scales contributing to the CMB constraint: $l_\text{min} = 2$, and $l_\text{max} \approx 2000$, as \Planck\ measurements are dominated by noise at higher $l$s \cite{Planck:2018vyg}.
Since the hierarchy of scales accessible to $\mu T$ cross-correlations is much greater than that available to the CMB ($T,E$)-bispectrum, for sufficiently large $\alpha$, the $\mu T$ measurement will always feature a larger signal-to-noise ratio.

Note that in the literature there are existing mechanisms that produce a similar faster-than-local growth in the squeezed limit of bispectrum. For example, modifying the initial state of the perturbations can lead to poles in the folded-limit ($k_2+k_3-k_1\to 0$ etc.)  that reduces to our $\alpha=1$ shape upon a further requirement $k_3\ll k_1$ \cite{Chen:2006nt,Holman:2007na,Ganc:2011dy,Ghosh:2022cny}. However, the squeezed-limit growth in these models originates from the UV dynamics of unknown excited states, and is limited to the range $k_S/k_L\lesssim 10^2$ by consistency \cite{Flauger:2013hra} (see also UV-complete models in \cite{Chen:2010bka}). In contrast, signals from the cosmological tachyon collider are generated from the IR physics of the Bunch-Davies vacuum state and can reach up to $k_S/k_L\lesssim 10^7$, which is more compatible with the hierarchy of scales in the $\mu T$ cross correlation. Other proposals include introducing sharp features in the inflaton potential \cite{Arroja:2011yu}, or evoking anisotropic inflation \cite{Dey:2013tfa}. However, we notice that scale invariance (and rotational invariance as well in the case of anisotropic inflation) is typically violated in the observable bispectrum. For comparison, our bispectrum is to a good approximation invariant under scale transformation and rotation. Note also that despite the mild level of resemblance of our setup to the well-known hybrid or waterfall inflation scenario \cite{Linde:1993cn,Dvali:1994ms,Copeland:1994vg}, the dynamical aspects are drastically different. Typical hybrid models contain a waterfall phase that lasts a few efolds at the very end of inflation, and produces primarily local non-Gaussianities \cite{Alabidi:2006wa,Byrnes:2008zy,Dutta:2008if,Mulryne:2011ni}. In contrast, our tachyon sits in the middle of inflation and can survive $\mathcal{O}(16)$ efolds, leading to a faster-than-local non-Gaussian signal.\\ 

The rest of this paper is structured as follows. In Section \ref{TachyondSSection} we start by studying the free theory of a tachyon in de Sitter spacetime. Then we present concrete models in Section \ref{ModelSection} and compute the non-Gaussian signal in Section \ref{CTCSignalSection}, before giving phenomenological templates in Section \ref{sec:TemplateSection} and forecasts on the $\mu T$-correlator from future CMB distortion measurements in Section \ref{muTProspectsSection}. We conclude with a discussion and an outlook in Section \ref{ConclusionsSect}. Throughout the paper we use the metric signature convention $(-,+,+,+)$ and set $c=\hbar=1$.


\section{A spectator tachyon in de Sitter}\label{TachyondSSection}

In this section we briefly discuss the free theory of a scalar field in de Sitter with a negative squared mass.

Let us start by considering a spectator scalar field $\chi$ in the Poincaré patch of de Sitter spacetime,
\begin{align}
	S_0[\chi]=\int \d^4x\sqrt{-g}\left[-\frac{1}{2}(\partial\chi)^2-\frac{1}{2}m^2_\chi \chi^2\right]~,
\end{align}
where the de Sitter metric is $g_{\mu\nu}=a^2(\eta)\eta_{\mu\nu}$ with $a(\eta)=-1/(H\eta)$. In conventional models, for example those focusing on the cosmological collider signatures, the scalar field $\chi$ has a positive squared mass, i.e. $m_\chi^2>0$, and the free theory is stable. The one-particle states of $\chi$ furnish a unitary irreducible representation of the dS group. However, for reasons mentioned in the introduction, the resulting signals are suppressed in the squeezed limit bispectrum. Instead, here we study the case of a \textit{negative} squared mass, namely
\begin{align}
	m^2_\chi=-\tilde{m}^2<0~.
\end{align}
Varying the action with respect to the field, we obtain the usual Klein-Gordon equation, which in momentum space reads
\begin{align}
	\left(\eta^2\frac{\partial}{\partial\eta^2}-2\eta\frac{\partial}{\partial\eta}+k^2\eta^2-\frac{\tilde{m}^2}{H^2}\right)\chi(\eta,k)=0~.\label{SpectatorTachyonEoM}
\end{align}
In the early-time limit $-k\eta\to \infty$, the mass term is negligible and the field admits a Minkowski asymptotic behaviour which allows us to impose the standard Bunch-Davies initial condition and select the adiabatic vacuum. The solution to \eqref{SpectatorTachyonEoM} is then given in terms of the Hankel function,
\begin{align}
	\chi(\eta,k)=f_{\tilde{\nu}}(\eta,k)\equiv -\frac{i\sqrt{\pi}}{2} e^{i\pi\left(\frac{\tilde\nu}{2}+\frac{1}{4}\right)} H (-\eta)^{3/2} H_{\tilde\nu}^{(1)}(-k\eta)~,\quad \tilde{\nu}\equiv\sqrt{\frac{9}{4}+\frac{\tilde{m}^2}{H^2}}~.\label{SpectatorTachyonModeFun}
\end{align}
Note that, compared to a stable scalar, the sole difference here is the definition of the dimensionless mass parameter $\tilde{\nu}$, which now takes values with $\tilde{\nu}>3/2$.\\

In flat spacetime, such fields with a negative squared mass are known as tachyons, due to the fact that the group velocity of short-wavelength excitations is superluminal in the UV,
\begin{align}
	v_{\rm g}=\frac{d \omega}{d k}=\frac{k}{\sqrt{k^2-\tilde{m}^2}}>1~,\quad k> \tilde{m}~,
\end{align}
where we have used the dispersion relation $\omega^2=k^2-\tilde{m}^2$ of the tachyon. However, there is no violation of causality because such superluminal modes cannot be excited by local sources \cite{Aharonov:1969vu}, meaning that the front velocity that dictates the speed of information propagation is always luminal \cite{Shore:2007um,Babichev:2007dw},
\begin{align}
	v_{\rm f}=\lim_{k\to \infty}\frac{\omega}{k}=1~.
\end{align}
The problem for tachyons arises instead in the IR as an instability of long-wavelength excitations,
\begin{align}
	\omega=\pm i\sqrt{\tilde{m}^2-k^2}~,\quad k< \tilde{m}~,
\end{align}
suggesting that these modes will grow exponentially in time, i.e. $\chi(t,k) \propto e^{+|\omega| t},~k<\tilde{m}$. Thus the system is unstable against long-wavelength perturbations and tends to be driven to a new phase.\\

Analogously in the de Sitter case, tachyons also exhibit an instability in the IR. This is most easily seen from the late-time expansion of the mode function \eqref{SpectatorTachyonModeFun},
\begin{align}
	\chi(\eta,k)=\sqrt{\frac{2\pi}{k^3} } H e^{i \pi  (\tilde\nu/2 +1/4)} \left[\frac{\cot \pi \tilde\nu -i}{\Gamma (1+\tilde\nu)}\left(-\frac{k\eta}{2} \right)^{3/2+\tilde\nu} -\frac{\Gamma (\tilde\nu )}{\pi}\left(-\frac{k\eta}{2}\right)^{3/2-\tilde\nu }\right]+\mathcal{O}\left((-k\eta)^{5/2\pm\tilde\nu}\right)~.
\end{align}
Due to the fact that $\tilde{\nu}>3/2$, the second term in the bracket (which corresponds to the negative-frequency mode for heavy fields in the principal series) is now an \textit{increasing} function of time and quickly dominates over the other terms,
\begin{align}
	\chi(\eta,k)\approx-\sqrt{\frac{2\pi}{k^3} } H e^{i \pi  (\tilde\nu/2 +1/4)}\frac{\Gamma (\tilde\nu )}{\pi}\left(-\frac{k\eta}{2}\right)^{3/2-\tilde\nu }\propto e^{-3Ht/2}e^{+\tilde{\nu}Ht}~,\quad -k\eta\ll 1~.\label{chiLateTimeScaling}
\end{align}
Note that the first decaying factor can be attributed to cosmic dilution and the second growing factor to the more prominent tachyonic growth. Unlike the Minkowski case, the growth rate is uniform for all comoving momenta, as a consequence of the scale invariance of the de Sitter background. In other words, all modes undergo the same tachyonic growth when their physical wavelength is stretched beyond the Compton scale. After canonical quantisation, the late-time two-point function of the tachyon is therefore a red-tilted spectrum that possesses more fluctuation power in the IR,
\begin{align}
	\ex{\chi(\eta,\mathbf{k})\chi(\eta,-\mathbf{k})}'&\approx\frac{H^2}{2k^3}\left(\frac{\Gamma (\nu )}{\sqrt{\pi}/2}\right)^2\left(-\frac{k\eta}{2}\right)^{3-2\tilde\nu }~.\label{chiLateTimePowerSpectrum}
\end{align}
The energy-momentum tensor of the $\chi$ sector can be estimated from late-time two-point function using the mass term as
\begin{align}
	\ex{T^\mu_{~\nu}(\chi)}_{\mathcal{S}}\sim\int_{|\mathbf{k}|\in \mathcal{S}}\tilde{m}^2\ex{\chi(\eta,\mathbf{k})\chi(\eta,-\mathbf{k})}'=\tilde{m}^2H^2\frac{\Gamma(\tilde{\nu})^2}{\pi^3(3-2\tilde{\nu})}\left(k_{S}^{3-2\tilde{\nu}}-k_{L}^{3-2\tilde{\nu}}\right)\left(-\frac{\eta}{2}\right)^{3-2\tilde{\nu}}~,\label{TmunuEstimate}
\end{align}
where $\mathcal{S}=\{\mathbf{k}:k_L<|\mathbf{k}|<k_S\}$ denotes a shell of modes between long and short scales $k_L,k_S$ in momentum space. From \eqref{TmunuEstimate}, we observe that the tachyonic instability manifests itself in both the late-time growth as $\eta\to 0$ and in the IR divergence in the momentum cutoff $k_L\to 0$. The latter can be understood as the fact that modes with longer wavelengths enter the tachyonic phase earlier and enjoy more growth at a fixed time. Such divergences, if left unchecked, inevitably backreact on the spacetime background and lead to the breakdown of the perturbative description. As a result, in any realistic cosmological model of a tachyon, we need to regularise the instability and the tachyonic phase is necessarily transient. In the next section, we will introduce two phenomenological models to regularise the tachyon and embed them consistently into an inflationary context.\\

Before moving to model building, we point out that the interplay between tachyons and de Sitter space or inflation is \textit{not} an unfamiliar topic. It is known that there exist consistent formulations of exceptional scalars in exact de Sitter space that take discrete values of negative squared masses \cite{Bros:2010wa}. These de Sitter tachyons enjoy a gauge symmetry and can be quantised in a fully local and covariant fashion \cite{Epstein:2014jaa,Bonifacio:2021mrf}. However, their existence hinges on \textit{exact} de Sitter invariance, and is therefore of only marginal relevance for inflation, where de Sitter boosts are always spontaneously broken. On the other hand, inflationary tachyons may in fact already exist: the current observation prefers a concave inflaton potential \cite{Planck:2018jri}.


\section{Transient tachyons during inflation}\label{ModelSection}

Let us start by considering a two-field system with an inflaton field $\phi$ and a spectator field $\chi$ that is weakly coupled to the inflaton,
\begin{align}
	S=\int \d^4x\sqrt{-g}\left[\frac{1}{2}M_p^2 R-\frac{1}{2}(\partial\phi)^2-V_{\rm sr}(\phi)\right]+S_{\chi}+S_{\rm int}~,
\end{align}
where $V_{\rm sr}(\phi)$ is a flat potential giving rise to a slow-roll background solution $\ex{\phi}=\phi_0(t)$ with $\dot{\phi}_0\approx \text{const}$ and $a(t)=e^{H t}, ~H\simeq \sqrt{V_{\rm sr}(\phi_0)/3M_p^2}\approx \text{const}$. We demand that the $\chi$ sector evolves around a trivial background $\ex{\chi}=0$, with an energy density subdominant to that of the inflaton so that the background dynamics of the system remains effectively a (quasi) single-field slow-roll inflation. In this sense $\chi$ is a spectator field. Transient tachyonic instability can then be introduced through non-trivial structures in the $\chi$ action. In the following subsections, we show that this can be achieved by breaking either scale invariance or locality in the $\chi$ sector.


\subsection{Model I: A prolonged phase transition}\label{ModelISubSect}

A simple realisation of tachyonic instability is a prolonged phase transition with effective time dependence in the mass of $\chi$. This undoubtedly breaks scale invariance in the $\chi$ sector. Consider the following action of the $\chi$ field,
\begin{align}
	S_\chi=\int \d^4x \sqrt{-g}\left[-\frac{1}{2}(\partial\chi)^2-V_\chi\right]~,\quad V_\chi\equiv \frac{1}{2}m_\chi^2(\eta)\chi^2+\frac{1}{2N}\left(\frac{\chi^2}{\chi_0^2}\right)^{N}\chi_0^4~,\label{actionModelI}
\end{align}
where $\chi_0\gg H,~N\gg 1$ and
\begin{align}\label{mchiofeta}
	m^2_\chi(\eta)=\left\{\begin{aligned}
	m^2~,&\quad \eta< \eta_i\\
	-\tilde{m}^2~,&\quad \eta\geq \eta_i
	\end{aligned}\right.
\end{align}
\begin{figure}[b!]
   \centering
   \includegraphics[width=0.75\textwidth]{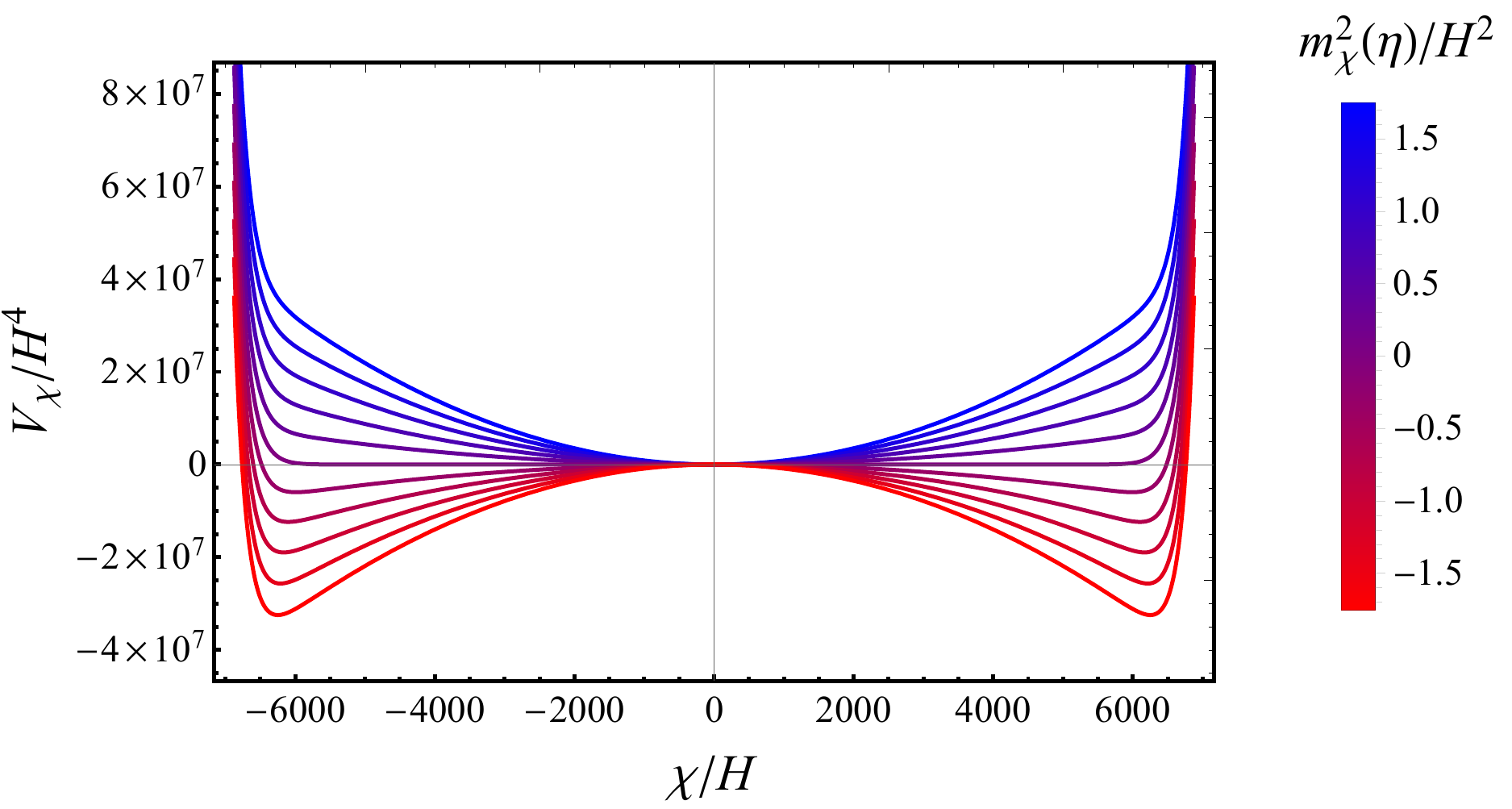}
   \caption{The time-dependent potential for the scalar field $\chi$ in Model I \eqref{actionModelI}. Depending on the sign of the mass parameter $m_\chi^2(\eta)$, the $\chi$ field changes from a stable field to an unstable field with tachyonic growth. Here the parameters are chosen as $\chi_0 = 10^4 H$ and $N=20$.}
   \label{ModelIPotential}
\end{figure}
is a time-dependent mass that triggers a (second-order) phase transition at $\eta_i$ (see Figure \ref{ModelIPotential} for an illustration of the $\chi$-field potential). Following the onset of the tachyonic phase, the one-point function of the $\chi$ field remains zero due to the $\mathbb{Z}_2$ symmetry of the action combined with the choice of initial state. The two-point function, however, grows exponentially in cosmic time following the discussion in the previous section. Such a growing variance of the fluctuations in the $\chi$ field is eventually halted by the positive power-law potential as it approaches $\ex{\chi^2}\sim \chi_f^2$, where it encounters the true minima lying at
\begin{align}
	\chi_f=\pm\left(\frac{\tilde{m}}{\chi_0}\right)^{\frac{1}{N-1}}\chi_0~.
\end{align}
Hence the exit time $\eta_f$ for the tachyonic phase can be estimated by the time when the variance of a mode $k=1/|\eta_i|$ reaches $\chi_f$. Using \eqref{chiLateTimePowerSpectrum} we find the condition
\begin{align}
	\frac{\eta_f}{\eta_i}=\left(\frac{\chi_f^2}{H^2}\right)^{\frac{1}{3-2\tilde{\nu}}}=\left[\frac{\chi_0^2}{H^2}\left(\frac{\tilde{m}}{\chi_0}\right)^{\frac{1}{N-1}}\right]^{\frac{1}{3-2\tilde{\nu}}}~.
\end{align}
In the limit $N\gg 1$, we have $\chi_f\approx \chi_0$ and the number of efolds of the tachyonic phase is approximately
\begin{align}
	\Delta \mathcal{N}_{fi}=-\ln \frac{\eta_f}{\eta_i}\approx \frac{1}{\tilde{\nu}-3/2}\ln \frac{\chi_0}{H}~,\quad N\gg 1~.\label{ApproxEfoldingsModelI}
\end{align}

Typical phase transitions last no longer than a few efolds in models of waterfall inflation and hybrid inflation \cite{Linde:1993cn,Dvali:1994ms,Copeland:1994vg}. In contrast, since we are interested in probing the tachyonic non-Gaussian signatures in the cross correlation of CMB $\mu$-distortions and temperature fluctuations, here we will require a \textit{prolonged} phase transition with 
\begin{align}
    \Delta \mathcal{N}_{fi}\gtrsim \ln 10^7 \sim 16\,,
\end{align} 
so that it spans from the CMB scales to the $\mu$-distortion scales\footnote{This prolonged phase transition does bear a sense of resemblance to the mild waterfall inflation scenario \cite{Clesse:2010iz,Kodama:2011vs,Abolhasani:2010kn}. However, our dynamical setups are different, leading to drastically different non-Gaussianity shapes, i.e. local (in mild-waterfall cases \cite{Abolhasani:2010kn,Clesse:2013jra}) vs beyond-local (in our case).}. After the tachyonic phase ends, the energy stored in the $\chi$ sector dissipates into thermalised $\chi$-particles around the new vacuum, which is then further diluted away during the remaining efolds of inflation.\\

Since we are mainly interested in the overall growth during the tachyonic phase, the smoothness of transition of the mass from negative to positive at $\eta_i$ is unimportant to leading order\footnote{The smoothness of the transition controls the adiabaticity of the perturbations and is tightly related to spontaneous particle production. In the case of sharp transitions, a typical $\mathcal{O}(1)$ number of particles are produced per Hubble patch.}. Thus, without loss of generality, we limit ourselves to the simplest case of a sharp transition. At early times, the linear theory for $\chi$ is that of a free massive field in de Sitter, whose mode function in momentum space and conformal time is (see \eqref{SpectatorTachyonModeFun} for the functional form of $f$)
\begin{align}
	\chi(\eta,k)=f_\nu(\eta,k)~,\quad \eta<\eta_i~,
\end{align}
where $\nu=\sqrt{9/4-m^2/H^2}<3/2$. After the triggering of the phase transition by an instantaneous change of mass at $\eta_i$, the mode function becomes a combination of positive and negative frequency solutions to the new equation of motion,
\begin{align}
	\chi(\eta,k)=A(\eta_i,k) f_{\tilde{\nu}}(\eta,k) + B(\eta_i,k) f_{\tilde{\nu}}^*(\eta,k)~,\quad \eta_i\leq\eta\leq \eta_f~,\label{chiMatchedModeFun}
\end{align}
where the coefficients $A(\eta_i,k),B(\eta_i,k)$ are found from the junction condition
\begin{align}
	\chi(\eta_i^-,k)&=\chi(\eta_i^+,k)~,\\
	\chi'(\eta_i^-,k)&=\chi'(\eta_i^+,k)~.
\end{align}
This junction condition preserves the Wronskian of the mode function,
\begin{align}
	\chi(\eta,k)\chi'^*(\eta,k)-\chi'(\eta,k)\chi^*(\eta,k)=i a^{-2}(\eta)~,\label{WronskianCondition}
\end{align}
as well as the canonical commutator relation after quantisation. Inserting \eqref{chiMatchedModeFun} into \eqref{WronskianCondition} yields a constraint on the Bogoliubov coefficients $A,B$,
\begin{align}
	|A|^2-|B|^2=1~,\label{BogoliubovConstraint}
\end{align}
For a sharp change of the mass parameter, the coefficients are naturally comparable for long-wavelength modes, i.e. $|B/A|=\mathcal{O}(1),\,-k\eta_i\ll 1$, whereas for short-wavelength modes, namely deep in the UV, the negative frequency component is suppressed by $|B/A|=\mathcal{O}(1/k^2\eta_i^{2}),\,-k\eta_i\gg 1$. It is useful to inspect the Pauli-Jordan function (equivalently, the field commutator) of the $\chi$ field in the late-time limit,
\begin{align}
	-i\ex{\, [\chi(\eta_1,\mathbf{k}),\,\chi(\eta_2,-\mathbf{k})]\, }'=-i\left[\chi(\eta_1,k)\chi^*(\eta_2,k)-\chi^*(\eta_1,k)\chi(\eta_2,k)\right]~.
\end{align}
Using the matched mode function \eqref{chiMatchedModeFun} and the constraint \eqref{BogoliubovConstraint}, one obtains
\begin{align}
	-i\ex{\, [\chi(\eta_1,\mathbf{k}),\,\chi(\eta_2,-\mathbf{k})]\, }'=\frac{H^2}{2\tilde{\nu}}(\eta_1\eta_2)^{3/2}\left[\left(\frac{\eta_1}{\eta_2}\right)^{\tilde\nu}-\left(\frac{\eta_2}{\eta_1}\right)^{\tilde\nu}\right]\left[1+\mathcal{O}(k^2\eta_1\eta_2)\right]~,\quad -k\eta_1,-k\eta_2\ll 1~.
\end{align}
Therefore, the Pauli-Jordan function decays as usual in the late time limit $\eta_1,\eta_2\to 0$ with the time ratio $\eta_1/\eta_2$ held fixed. This is again a manifestation of causality, namely super-horizon modes do not causally propagate. Comparing the Pauli-Jordan function to the power spectrum in \eqref{chiLateTimePowerSpectrum}, we find that $\chi$ field operators effectively commute in the late-time limit. This leads to a tremendously simplified description of the late-time tachyonic dynamics. In particular, all of the four Schwinger-Keldysh propagators of $\chi$ become identical and factorised,
\begin{align}
	&\nonumber\langle\chi_+(\eta_1,\mathbf{k})\chi_+(\eta_2,-\mathbf{k})\rangle'\approx\langle\chi_-(\eta_1,\mathbf{k})\chi_+(\eta_2,-\mathbf{k})\rangle'\approx\langle\chi_-(\eta_1,\mathbf{k})\chi_-(\eta_2,-\mathbf{k})\rangle'\approx\langle\chi_+(\eta_1,\mathbf{k})\chi_-(\eta_2,-\mathbf{k})\rangle'\\
	&\approx\frac{H^2}{2k^3}\left(\frac{\Gamma (\tilde\nu )}{\sqrt{\pi}/2}\right)^2\left|A e^{ i \pi  (2\tilde\nu +1)/4} +B e^{-i \pi  (2\tilde\nu +1)/4}\right|^2\left(-\frac{k\eta_1}{2}\right)^{\frac{3}{2}-\tilde\nu } \left(-\frac{k\eta_2}{2}\right)^{\frac{3}{2}-\tilde\nu }~,\quad -k\eta_1,-k\eta_2\ll 1~.\label{SKpropsFac}
\end{align}
We will see later in Section \ref{CTCSignalSection} that this factorisation property of the tachyon propagators gives rise to a neat characteristic template for the cosmological tachyon collider signal.

\paragraph{Origin of the time-dependent mass} Until now, we have assumed an explicit time dependence of the $\chi$ mass term $m_\chi^2(\eta)$, which serves as a trigger of a phase transition. However, one can embed this into a model where the approximate time-translation symmetry is spontaneously broken. For example, one can surmise a local feature in the scalar manifold,
\begin{align}
	m^2_\chi(\phi)=m^2-(m^2+\tilde{m}^2)\theta(\phi-\phi_i)~,
\end{align}
where $\theta(x)$ denotes the Heaviside step function that models a sharp change of the mass parameter, and $\phi_i$ parametrises the location of the feature. After the inflaton acquires a background $\phi_0(\eta)$ and spontaneously breaks time-translation symmetry, the beginning of the tachyonic phase is automatically set by $\phi_0(\eta_i)=\phi_i$. Notice that the presence of the feature does explicitly break the shift symmetry of the inflaton, and thereby introduces a direct coupling between the field perturbations via
\begin{align}
	m^2_\chi(\phi)\chi^2\supset m^2_\chi(\eta)\chi^2+\left.\frac{\partial m^2_\chi}{\partial\phi}\right|_{\phi=\phi_0(\eta)}\varphi \chi^2+\mathcal{O}(\varphi^2\chi^2)~,
\end{align}
where we have separated the full inflaton field into a classical background and its quantum fluctuations, i.e. $\phi=\phi_0+\varphi$. Direct and relevant couplings of the inflaton are usually dangerous since they can induce IR-divergent tadpole corrections to the inflaton background, threatening the perturbative control based on a preset background $\phi_0$. However, in our case the feature is localised in field space. Hence it gives rise to a coupling that is localised in time. Namely, the direct coupling is only instantaneous,
\begin{align}
	\left.\frac{\partial m^2_\chi}{\partial\phi}\right|_{\phi=\phi_0(\eta)}=-(m^2+\tilde{m}^2)\,\delta(\phi_0(\eta)-\phi_i)=\frac{(m^2+\tilde{m}^2)H}{\dot\phi_0}\,\eta_i\delta(\eta-\eta_i)~.
\end{align}
Furthermore, at $\eta_i$ the $\chi$ field has not yet been amplified by the tachyonic instability, thus its typical fluctuations are of order Hubble. One can thus estimate the typical size of the induced coupling in the Lagrangian as $(m^2+\tilde{m}^2)H/\dot\phi_0 \,\varphi\chi^2\sim H^6/\dot\phi_0\ll H^4$. Hence we do not expect large backreaction to the inflaton sector from this phenomenon. Note also that the exit from the tachyonic phase at $\eta_f$ is not controlled by the inflaton, but by the thermalisation process within the $\chi$ sector itself. Therefore, even at late times we do not expect any induced coupling to the inflaton, and hence no large backreaction either.

An alternative realisation of the time dependent mass in \eqref{mchiofeta} consists of introducing another rolling scalar field $\psi(\eta)$ with $\dot{\psi}\ll \dot{\phi}$. The essential ingredients are the same as in the inflaton case, however there is more freedom in choosing the dynamics. In the following we will simply assume that the mass becomes negative for an interval of time and our phenomenological analysis will be largely independent of the explicit model building realization.


\subsection{Model II: A high-pass tachyonic mass filter}

One might be concerned with the breaking of scale invariance in the previous model, even though, as we shall see later in Section \ref{CTCSignalSection}, it does not propagate to observables to leading order. Thus, it would be desirable to find a way to regulate the tachyonic instability while keeping manifest scale invariance. Here we present a second setup in which scale invariance is maintained, but at the cost of breaking locality in the unobservable $\chi$ sector. In particular, we consider a non-local free theory described by
\begin{align}
	S_\chi=\int \d^4x \sqrt{-g}\left[-\frac{1}{2}(\partial\chi)^2-\frac{1}{2}\chi \,m_\chi^2({\bm\nabla}^2)\chi\right]~,
\end{align}
where $m_\chi^2({\bm\nabla}^2)$ is a scale-dependent mass term that formally takes the form
\begin{align}
	m_\chi^2({\bm\nabla}^2)= m^2-(m^2+\tilde{m}^2)\frac{{\bm\nabla}^{2N}}{{\bm\nabla}^{2N}+\delta^{2N}}~,
\end{align}
with $\delta\ll H$ and $N\gg 1$. Here ${\bm\nabla}^2\equiv \partial_i^2/a^2(\eta)$ is the spatial Laplacian operator. The mass operator $m_\chi^2({\bm\nabla}^2)$ is apparently non-local as it involves an inverted differential operator. Thus it is better defined in momentum space as a high-pass filter,
\begin{align}
	m_\chi^2(k^2/a^2)= m^2-(m^2+\tilde{m}^2)\frac{(k/a)^{2N}}{(k/a)^{2N}+\delta^{2N}}~.
\end{align}
Note that such non-locality is mild as it is analytic around zero momentum $k/a\to 0$ (it does not involve any inverse Laplacian by itself). Therefore the above non-locality does not mediate any long-range interactions, but rather resembles more the finite-ranged Yukawa forces in nuclear physics\footnote{In fact, in the opposite regime where $N=1$ and $\delta>H$, this filtered-mass operator could come from, for example, integrating out heavy degrees of freedom when there is a sound speed hierarchy \cite{Jazayeri:2023kji}.}. In fact, for $N\gg 1$, the only energy scales that exhibit non-locality are those around the threshold scale $\delta$. Away from the threshold scale, $\chi$ behaves as two copies of a local free theory with different masses,
\begin{align}
	m^2_\chi(k^2/a^2)=\left\{\begin{aligned}
	m^2~,&\quad k/a(\eta)\ll \delta\\
	-\tilde{m}^2~,&\quad k/a(\eta)\gg \delta
	\end{aligned}\right.~.
\end{align}
Therefore, for a given mode labelled by its comoving momentum $k$, the filtered mass is initially tachyonic and the mode experiences exponential growth after becoming non-relativistic at $k/a\sim \tilde{m}$. The growth eventually stops when the energy scale of the mode passes the threshold $k/a\sim\delta$, where the mass turns positive again, regularising the instability. The duration of the tachyonic phase for any given mode is thus
\begin{align}
	\Delta\mathcal{N}_{fi}\approx\ln\frac{\tilde{m}}{\delta}~.
\end{align}
Since the amount of growth experienced by different modes is the same for all $k$, scale invariance is automatically guaranteed.

The analysis of the free theory mode function is almost identical to Model I, with the Bogoliubov coefficients in \eqref{chiMatchedModeFun} and \eqref{SKpropsFac} now simplified to $A=1,B=0$, as there is no transition before the onset of the tachyonic phase.


\section{The cosmological tachyon collider}\label{CTCSignalSection}

Having understood the background and linear dynamics of a transient tachyonic phase, we now move on to the non-linear dynamics and compute observables. More specifically, we will focus on the bispectrum of curvature perturbations arising from a tachyon exchange process as depicted below.
\begin{center}
\begin{dSdiagram}[aspect=0.15, width=5cm]
   \node(L) at (-0.25*\dSwidth,-0.25 * \dSwidth) [vertex,label={left:$\eta_1$}];
   \node(R) at (0.15*\dSwidth, -0.5 * \dSwidth) [vertex,label=right:$\eta_2$];
   \coordinate[label=above:$\zeta(\vb k_1)$](k1) at (-.35*\dSwidth,0);
   \coordinate[label=above:$\zeta(\vb k_2)$](k2) at (-.15*\dSwidth,0);
   \coordinate[label=above:$\zeta(\vb k_3)$](k3) at (.35*\dSwidth,0);
   \draw (k1) -- (L);
   \draw (k2) -- (L);
   \draw [dashed] (L) -- (R);
   \draw (R) node[vertex] -- (k3);
\end{dSdiagram}
\end{center}
We will assume that all three Fourier modes in the bispectrum exit the horizon \textit{during} the tachyonic phase, i.e. $(-\eta_i)^{-1}<k_1,k_2,k_3<(-\eta_f)^{-1}$. The squeezed limit of this bispectrum is of particular interest, since for IR-convergent couplings, we expect that the typical interaction time is dictated by the horizon-exit time of the shortest mode participating in the interaction. Thus in the squeezed limit $k_1,k_2\gg k_3$, the characteristic interaction times of the triple vertex ($\eta_{1,*}\sim -k_{12}^{-1}$)\footnote{Throughout the paper we use the abbreviation $k_{ij}\equiv k_i+k_j=|\mathbf{k}_i|+|\mathbf{k}_j|$.} and the quadratic mixing ($\eta_{2,*}\sim -k_3^{-1}$) forms a hierarchy, i.e. $\eta_{2,*}/\eta_{1,*}\sim k_{12}/k_3\gg 1$. Due to the tachyonic growth of the exchanged $\chi$ mode during its propagation from $\eta_{2,*}$ to $\eta_{1,*}$, the curvature bispectrum is amplified correspondingly (see e.g. \eqref{chiLateTimeScaling}),
\begin{align}
	B_3^{\zeta}(k_1,k_2,k_3)\equiv \ex{\zeta(k_1)\zeta(k_2)\zeta(k_3)}'\propto \left(\frac{\eta_{1,*}}{\eta_{2,*}}\right)^{3/2-\tilde{\nu}}\sim \left(\frac{k_{12}}{k_3}\right)^{\tilde{\nu}-3/2}~.\label{ScalingIntuition}
\end{align}
Since $\tilde{\nu}>3/2$, the bispectrum is expected to grow in the squeezed limit. This provides an example of an enhanced UV-IR mixing with an \textit{increasing} correlation between scales that are more separated. Of course, such a counter-intuitive growth is eventually cut off by the end of the tachyonic phase, in a way that either breaks (Model I) or preserves (Model II) scale invariance.

Most of the discussion in the rest of this section is agnostic about the model realisations of a transient tachyon. We only rely on the most salient feature of the system, namely the late-time growth due to a negative squared mass. The model dependence only kicks in later, when we discuss the extreme IR.


\subsection{Interactions and IR convergence}
We will proceed by introducing the coupling of the transient tachyon to the inflaton. Without loss of generality, let us first focus on Model I. Working at the level of inflationary perturbations in the spatially flat gauge, we consider a family of EFT operators that respect the unbroken spacetime symmetries (dilation, translations and rotations) as well as the shift symmetry of the inflaton,
\begin{align}
	\nonumber S_{\text{int}}=\int \d^4x\sqrt{-g}\Bigg[&\sum_{m,n} \rho_{mn}\, {\bm \nabla}^{2m}(-H\eta\partial_\eta)^n\varphi\\
	&+ \sum_{m,n,p,q,l}\lambda_{mnpql} \,{\bm \nabla}^{2m+l}(-H\eta\partial_\eta)^n\varphi\cdot {\bm \nabla}^{2p+l}(-H\eta\partial_\eta)^q\varphi +\mathcal{O}(\varphi^3)\Bigg]\chi +\mathcal{O}(\chi^2)~,\label{InteractionEFTLagrangian}
\end{align}
where we have omitted terms that do not contribute to the bispectrum at leading order at tree level. We have also applied integration-by-parts to move all the derivatives from $\chi$ to $\varphi$\footnote{This process generates boundary terms which we assume to vanish by formally compactifying space and imposing the decay of the mass-restored $\chi$ field at the future boundary.}. Note that some of the operators (e.g. the $\rho_{01}$ term) in \eqref{InteractionEFTLagrangian} are IR-divergent due to the anomalous late-time growth of the tachyon. These IR divergences reflect the fact that the copious production of $\chi$ modes sources strong couplings to the inflaton and hence to the curvature fluctuations even when they are outside the horizon. Since our intuition in \eqref{ScalingIntuition} rests upon the existence of a characteristic interaction time at horizon exit, having IR divergent interactions would invalidate this picture, leading to the loss of the tachyonic signal that we seek. In order to keep the theory weakly coupled, we need to make a restriction on the IR behaviour of the set of operators in $S_{\text{int}}$. Namely, we require that all IR-divergent operators have a vanishing Wilson coefficient. For the operators explicitly shown in \eqref{InteractionEFTLagrangian}, we can make use of the IR scalings
\begin{align}
	\varphi(\eta) \sim (-\eta)^0~,\quad -\eta\partial_\eta \varphi(\eta) \sim (-\eta)^2~,\quad \chi(\eta) \sim (-\eta)^{3/2-\tilde{\nu}}~,\quad \eta\to 0~,
\end{align}
and write
\begin{align}
	S_{\text{int}}\sim\int \d\eta\left[\sum_{m,n} \rho_{mn} (-\eta)^{2(m-\delta_{n0})-1/2-\tilde{\nu}}+ \sum_{m,n,p,q,l}\lambda_{mnpql} (-\eta)^{2(m+p+l-\delta_{n0}-\delta_{q0})+3/2-\tilde{\nu}} \right]+\cdots~.
\end{align}
Convergence at $\eta\to 0$ requires (the integer $m$ here should not be confused with the mass)
\begin{align}
	\text{IR-finiteness constraints:}\quad\left\{
	\begin{aligned}
	\rho_{mn}&=0~,\quad 0\leq m<\frac{1}{2}\left(\tilde{\nu}-\frac{1}{2}\right)+\delta_{n0}\\
		\lambda_{mnpql}&=0~,\quad 0\leq m+p+l<\frac{1}{2}\left(\tilde{\nu}-\frac{5}{2}\right)+\delta_{n0}+\delta_{q0}\\
		&\cdots
	\end{aligned}\right.~.\label{IREFTConstraint}
\end{align}
Analogously, we set to zero all other Wilson coefficients of IR-divergent operators in the omitted terms of the whole EFT tower in $S_{\text{int}}$. In this way, we obtain a consistent weakly coupled theory without IR divergences. Crucially, the selection criterion \eqref{IREFTConstraint} is compatible with the renormalisation of Wilson coefficients by loop diagrams, since IR-finite operators \textit{cannot} induce IR-divergent operators via loops, otherwise there would be a mismatch in the late-time behaviour of correlators in the loop calculation and the effective tree calculation. For models that are relevant for phenomenology, the tachyon mass is typically between $3/2<\tilde{\nu}<5/2$, and the lowest-order interactions are thus of dimension-5:
\begin{align}
	S_{\text{int}}=\int \d\eta \d^3x\,\left[\rho\, a(\eta)\partial_i^2\varphi'\chi+\lambda\, a^2(\eta) \varphi'^2\chi\right]+\cdots~,\label{ReducedInteractionLagrangian}
\end{align}
where we have abbreviated $\rho\equiv \rho_{11},\lambda\equiv \lambda_{01010}$. Finally we comment that these two operators can also be considered as descendants of building blocks in the EFT of Inflation (EFTI) \cite{Cheung:2007st}, for example,
\begin{align}
	(n^\mu\nabla_\mu\delta K)\chi&\supset \partial_i^2\varphi'\chi~,\\
	(\delta g^{00})^2\chi&\supset \varphi'^2\chi~.
\end{align}
However, one should be cautious when rewriting (at least a subset of) IR-finite operators into combinations of the EFTI operators, since there can be intricate cancellations among them.


\subsection{Shapes of the bispectrum}
The non-Gaussianity arising from the leading couplings \eqref{ReducedInteractionLagrangian} can be straightforwardly computed using the Schwinger-Keldysh formalism (see \cite{Chen:2017ryl} for a review). In the weak mixing limit, $\rho H<1$, the leading diagram reads
\begin{align}
	\nonumber B_3^\varphi(k_1,k_2,k_3)=2\lambda\rho\sum_{\sf a b=\pm}{\sf a b}\int \d\eta_1 \d\eta_2 a^2(\eta_1) a(\eta_2)& \partial_{\eta_1} \mathcal{K}_{\sf a}^{\varphi}(\eta_1,k_1) \partial_{\eta_1} \mathcal{K}_{\sf a}^{\varphi}(\eta_1,k_2)\\
	&\mathcal{G}_{\sf a b}^\chi (\eta_1,\eta_2,k_3)\, k_3^2\, \partial_{\eta_2} \mathcal{K}_{\sf a}^{\varphi}(\eta_2,k_3)+\text{2 perms}~,
\end{align}
where $\mathcal{K}^X$ and $\mathcal{G}^X$ are the free-theory bulk-boundary and the bulk-bulk propagators for the field $X=\varphi,\chi$. As previously mentioned in Section \ref{ModelISubSect}, the bulk-bulk propagator of the tachyon enjoys a great simplification due to the decay of the Pauli-Jordan function at late times: it factorises to leading order. Since we are working in the squeezed limit $k_3\ll k_1,k_2$, the cubic vertex receives the largest contribution from the horizon-exit time ($\eta_{1,*}\sim -k_{12}^{-1}$) of the hard inflatons, which translates to a late time ($-k_3 \eta_{1,*}\ll 1$) for the soft tachyon. On the other hand, the quadratic mixing vertex receives the largest contribution from the horizon-exit of both the soft inflaton and the soft tachyon ($\eta_{2,*}\sim -k_3^{-1}$), for which the factorisation is marginally valid. Therefore we can adopt the factorisation property \eqref{SKpropsFac} as a good approximation to compute the leading signal\footnote{A similar late-time expansion technique has been applied to the computation of oscillatory cosmological collider signals \cite{Chua:2018dqh}, where a mismatch of mass prefactor in the signal has been spotted and explained in \cite{Wang:2020ioa,Tong:2021wai}. In our case, the tachyon mass is of order the Hubble scale, thus we expect at most an $\mathcal{O}(\nu)\sim \mathcal{O}(1)$ prefactor uncertainty in the signal amplitude.}. We also discard the contribution before the phase transition as well as that after the restoration of a positive squared mass, as they are subdominant in the time integral as long as all three modes exit the horizon during the bulk of tachyonic phase (this also suggests $|A(\eta_i,k_3)|\approx 1, |B(\eta_i,k_3)|\approx 0$). Inserting the mode functions yields
\begin{align}
	\nonumber B_3^\varphi(k_1,k_2,k_3)\approx& \frac{H^5\lambda\rho}{4\pi} \frac{4^{\tilde{\nu}} \Gamma(\tilde{\nu})^2}{k_1 k_2 k_3^{-1+2\tilde{\nu}}}\int_{-\infty}^{0} d \eta_1 (-\eta_1)^{3/2-\tilde{\nu}}\sin(k_{12}\eta_1)\times\int_{-\infty}^{0} d\eta_2 (-\eta_2)^{3/2-\tilde{\nu}}\sin(k_3\eta_2)\\
	&+\text{2 perms}~.
\end{align}
Performing the factorised time integrals, we obtain in the squeezed limit $k_3\ll k_1,k_2$,
\begin{align}
	B_3^\varphi(k_1,k_2,k_3)\approx& H^5\lambda\rho\,\mathcal{C}(\tilde{\nu})\frac{2}{k_1 k_2 k_{12} k_3^3}\left(\frac{k_{12}}{2k_3}\right)^{\tilde{\nu}-3/2}+\text{2 perms}~,\label{B3phiApprox}
\end{align}
where the mass-dependent coefficient is given by
\begin{align}
	\mathcal{C}(\tilde{\nu})\equiv \frac{2^{3(\tilde{\nu}-3/2)}}{\pi}\Gamma(\tilde{\nu})^2 \Gamma\left(\frac{5}{2}-\tilde{\nu}\right)^2\cos^2\left[ (2\tilde\nu +1)\pi/4\right]~.
\end{align}
The primordial curvature bispectrum can then be obtained by converting the spatially-flat $\varphi$-gauge to the homogeneous-inflaton $\zeta$-gauge using $\zeta=-(H/\dot\phi_0)\varphi$,
\begin{align}
	B_3^\zeta(k_1,k_2,k_3)\approx(2\pi^2\Delta_\zeta^2)^2\, f_{\rm NL}^{(\alpha)} \,\left[\frac{2}{k_1 k_2 k_{12} k_3^3}\left(\frac{k_{12}}{2k_3}\right)^{\alpha}+\text{2 perms}\right]~,\quad \alpha=\tilde{\nu}-3/2>0~,\label{B3zetaApprox}
\end{align}
where the overall amplitude is given by the dimensionless parameter
\begin{align}
	f_{\rm NL}^{(\alpha)}=-\frac{2}{\pi \Delta_\zeta} H^2\lambda\rho\,\mathcal{C}(\alpha+3/2)~.\label{fNLDef}
\end{align}
Notice that our definition of $\fnlA$ is somewhat different from the standard one in that the eventually large factor $[k_{12}/(2k_3)]^\alpha$ is not part of the definition of $\fnlA$. Because of this, the correct rough estimate for the size of the signal is $\fnlA [k_{12}/(2k_3)]^\alpha$. This fact is plainly visible for example in our perturbativity bounds, \eqref{pertbound1} and \eqref{pertbound2}.

Inspecting the leading-order squeezed limit with $k_1=k_2=k_S$ and $k_3=k_L$, we find the characteristic shape pattern
\begin{align}
	B_3^\zeta(k_S,k_S,k_L)\xrightarrow{k_S\gg k_L} (2\pi^2\Delta_\zeta^2)^2 \,f_{\rm NL}^{(\alpha)} \,\frac{1}{k_S^3 k_L^3}\left(\frac{k_S}{k_L}\right)^\alpha~.\label{SqueezedLimitBispectrum}
\end{align}
Hence in agreement with our physical intuition, squeezing the momentum triangle effectively increases the time lapse between the left and right vertices, which allows for more growth during the propagation of the tachyon, and in turn an increasing correlation between separated scales. Note that the well-known local-shape non-Gaussianity\footnote{Such a signal cannot appear in $\mu T$ in single-field attractor inflation because of Maldacena's consistency relation \cite{Maldacena:2002vr} and the absence of projection effects in $\mu$ \cite{Cabass:2018jgj}.} corresponds to the case $\alpha=0$, upon the identification of the amplitude parameter
\begin{align}
    f_{\rm NL}^{(0)}=\frac{6}{5}f_{\rm NL}^{\rm loc}~,
\end{align}
where the convention for the normalisation of local non-Gaussianity follows from \cite{Planck:2019kim}. However, for $\alpha>0$, the growth of the signal in the squeezed limit is faster than for local non-Gaussianity.

\paragraph{Scale dependence and the termination of the signal growth} Physically one expects that such a UV-IR correlation must cease to grow at some point. The attentive reader may have already noticed that, despite the explicit breaking of scale invariance in the $\chi$ sector in Model I, the leading bispectrum \eqref{B3zetaApprox} turns out to be scale invariant. This follows from two facts: we considered IR-finite interactions and we assumed that all three modes exited the horizon during the tachyonic phase. IR convergence ensures that the time when a correlation is imprinted depends on the momentum of a given mode rather than on boundaries $\{\eta_i,\eta_f\}$ of the tachyonic phase. The assumption that the momenta lie within the tachyonic phase implies that the mode functions and the free-theory propagators are scale covariant and independent of $\{\eta_i,\eta_f\}$. Had we considered modes that are outside the tachyonic window, the dependence on $\{\eta_i,\eta_f\}$ would have re-appeared and would have engendered a breaking of scale invariance. For example, suppose we kept $k_L=k_3$ fixed, and increased $k_S=k_1=k_2$ such that $k_S>(-\eta_f)^{-1}$. The mode function of $\chi(\eta,k_L)$ would change to a decreasing function of time because a positive squared mass has been restored. The outcome is therefore again a correlator that decreases with respect to increasing $k_S$. 

The story for Model II is similar, but the transition point is given in terms of the scale ratio rather than the short scale alone. To summarise, the asymptotic behaviour of the our signal is more properly written as
\begin{align}
	B_3^\zeta(k_S,k_S,k_L)\xrightarrow{k_S\gg k_L}(2\pi^2\Delta_\zeta^2)^2 \,f_{\rm NL}^{(\alpha)} \,\frac{1}{k_S^3 k_L^3}\left(\frac{k_S}{k_L}\right)^\alpha\times \mathcal{W}_\beta(k_S,k_L,\eta_f)~,\label{CorrectedBispectrum}
\end{align}
where $\mathcal{W}_\beta$ is an envelope function that restores UV-IR decoupling in the limit $k_S\to \infty$. For our models, one viable parametrisation is a broken power law:
\begin{align}
	\mathcal{W}_\beta(k_S,k_L,\eta_f)=\left\{\begin{aligned}
	~\Bigg[1+\left(-k_S\eta_f\right)^{\alpha-\beta}\Bigg]^{-1}~,&\quad \text{Model I}\\
	~\left[1+\left(\frac{\delta}{\tilde{m}}\frac{k_S}{k_L}\right)^{\alpha-\beta}\right]^{-1}~,&\quad \text{Model II}
	\end{aligned}\right.~,\quad\beta\equiv \nu-3/2<0~.
\end{align}
It is easy to see that within the tachyonic window where $k_S<(-\eta_f)^{-1}$ or $k_S/k_L<\tilde{m}/\delta$, \eqref{CorrectedBispectrum} reduces to \eqref{SqueezedLimitBispectrum}. For even shorter $k_S$ modes living outside the tachyonic window, the scaling exponent is changed back to $\beta<0$, which yields a correlation function now decreasing in the squeezed limit, as in conventional models.


\subsection{Constraints and parameter space} \label{sec:parameter constraints}

The faster-than-local squeezed signal of the cosmological tachyon collider is best probed by correlating modes with a vast separation of scales. Given the relatively pristine CMB data set, one possibility is to examine the cross correlation between the CMB $\mu$-type distortions and temperature anisotropies. To compare to data we need to take into account a series of constraints on the self-consistency of the model. In addition, the signal amplitude $f_{\rm NL}^{(\alpha)}$ is proportional to the mixing and triple couplings, which are also bounded from above by the consistency of perturbation theory and current observations. In this subsection we list the consistency constraints on the background model and on the couplings. 

Throughout the discussions below, the scale hierarchy between the CMB $\mu$-type distortions and temperature anisotropies plays an important role. We shall take the scale of the $l=2$ mode with $k_{\rm CMB}=2/r_{\rm LS}\simeq 1.4\times 10^{-4}\,\text{Mpc}^{-1}$ to be the typical scale of CMB temperature anisotropies, where $r_{\rm LS}=14\, \text{Gpc}$ is the distance to the last scattering surface. A pessimistic estimate of the $\mu$-type distortion scale is $k_\mu=10^4\,\text{Mpc}^{-1}$ which leads to $\log_{10}(k_\mu/k_{\rm CMB})\simeq 7.8$. However, due to the smearing effect of the window function \cite{Cyr:2023pgw}, this may overestimate the observable $\mu$ modes. An optimistic estimate is instead $k_\mu=740\,\text{Mpc}^{-1}$ \cite{CMB-S4:2023zem}, which leads to $\log_{10}(k_\mu/k_{\rm CMB})\simeq 6.7$. Considering the uncertainties in the estimates, we shall take $\log_{10}(k_\mu/k_{\rm CMB})\simeq 7.0$ as a simple benchmark value for the scale hierarchy used in consistency checks.
\begin{itemize}
    \item \textit{Sustaining a successful inflationary background}. The total energy stored in the $\chi$ sector at the end of the tachyonic phase must be subdominant compared to the vacuum energy of the inflaton,
    \begin{align}
        \ex{T^0_{~0}(\chi)}\ll\ex{T^0_{~0}(\phi_0)}~.
    \end{align}
    In Model I, using \eqref{TmunuEstimate}, we can write
    \begin{align}
        \tilde{m}^2 H^2\frac{\Gamma(\tilde{\nu})^2}{\pi^3(2\tilde{\nu}-3)}\left(\frac{\eta_f}{\eta_i}\right)^{3-2\tilde{\nu}} \ll 3 M_p^2 H^2\,.
    \end{align}
    For $3/2<\tilde\nu<5/2$, this constraint simply translates to
    \begin{align}
        \text{Model I :}\quad\chi_0\ll M_p
    \end{align}
    after applying \eqref{ApproxEfoldingsModelI}. In Model II, the effective growth efolds are determined by the ratio $\tilde{m}/\delta$, and we have instead
    \begin{align}
        \tilde{m}^2 H^2\frac{\Gamma(\tilde{\nu})^2}{\pi^3(2\tilde{\nu}-3)}\left(\frac{\delta}{\tilde m}\right)^{3-2\tilde{\nu}} \ll 3 M_p^2 H^2~.
    \end{align}
    which reduces to
    \begin{align}
        \text{Model II :}\quad\left(\frac{\delta}{H}\right)^{\nu-3/2}\gg \frac{H}{M_p}\,.
    \end{align}
    for $3/2<\tilde\nu<5/2$.

    \item \textit{A sufficiently wide tachyonic window of scales}. Since we are probing the very squeezed limit via the $\mu T$-correlation, the scales $k_S\sim k_{\mu}$ and $k_L\sim k_{\rm CMB}$ must lie in the bulk of the tachyonic window,
    \begin{align}
        e^{\Delta\mathcal{N}_{fi}}>\frac{k_{\mu}}{k_{\rm CMB}}\simeq 10^7~,
    \end{align}
    which translates to
    \begin{align}
        \text{Model I :}&\quad\left(\frac{\chi_0}{H}\right)^{1/\alpha}>10^7~,\\
        \text{Model II :}&\quad \frac{\tilde m}{\delta}>10^7~.
    \end{align}

    \item \textit{Perturbativity}. The inflaton propagator is corrected to leading order by (i) the tachyon exchange diagram due to the mixing coupling, and (ii) the self-energy diagram due to the triple coupling. To ensure perturbativity and the consistent use of free-theory propagators, we therefore require
    \begin{align}
       \begin{dSdiagram}[aspect=0.15, width=2.8cm]
   \node(L) at (-0.2*\dSwidth,-0.3 * \dSwidth) [vertex];
   \node(R) at (0.2*\dSwidth, -0.3 * \dSwidth) [vertex];
   \coordinate(lc) at (-.32*\dSwidth,-0.15 * \dSwidth);
   \coordinate(rc) at (.32*\dSwidth,-0.15 * \dSwidth);
   \coordinate(k1) at (-.35*\dSwidth,0);
   \coordinate(k2) at (.35*\dSwidth,0);
   \draw (k1) .. controls (lc) ..  (L);
   \draw [dashed] (L) -- (R);
   \draw (R) .. controls (rc) .. (k2);
\end{dSdiagram}
   \ , \ 
   \begin{dSdiagram}[aspect=0.15, width=2.8cm]
   \node(L) at (-0.2*\dSwidth,-0.3 * \dSwidth) [vertex];
   \node(R) at (0.2*\dSwidth, -0.3 * \dSwidth) [vertex];
   \coordinate(lc) at (-.32*\dSwidth,-0.15 * \dSwidth);
   \coordinate(rc) at (.32*\dSwidth,-0.15 * \dSwidth);
   \coordinate(k1) at (-.35*\dSwidth,0);
   \coordinate(k2) at (.35*\dSwidth,0);
   \draw (k1) .. controls (lc) ..  (L);
   \draw[dashed] (R) arc [start angle=0, end angle=180, radius=.2*\dSwidth];
   \draw (R) arc [start angle=0, end angle=-180, radius=.2*\dSwidth];
   \draw (R) .. controls (rc) .. (k2);
\end{dSdiagram}
\quad &< \quad
\begin{dSdiagram}[aspect=0.15, width=2.8cm]
   \coordinate(L) at (-0.2*\dSwidth,-0.45 * \dSwidth) ;
   \coordinate(R) at (0.2*\dSwidth, -0.45 * \dSwidth) ;
   \coordinate(k1) at (-.35*\dSwidth,0);
   \coordinate(k2) at (.35*\dSwidth,0);
   \draw (k1) .. controls (L) and (R) .. (k2);
\end{dSdiagram}
    \end{align}
    Using the approximate factorisability \eqref{SKpropsFac}, this reduces to a bound on the size of couplings,
    \begin{align}
        \left(\rho H\right)^2 2^{2-\alpha} \mathcal{C}(\alpha+3/2)&< 1~,\label{MixingPertConstraint}\\
        \frac{1}{(4\pi)^2}\left(\lambda H\right)^2& < 1~.\label{phiPropPertConstraint}
    \end{align}
    The perturbativity constraint from the renormalisation of the tachyon propagator given by
    \begin{align}
    \begin{dSdiagram}[aspect=0.15, width=2.8cm]
   \node(L) at (-0.2*\dSwidth,-0.3 * \dSwidth) [vertex];
   \node(R) at (0.2*\dSwidth, -0.3 * \dSwidth) [vertex];
   \coordinate(lc) at (-.32*\dSwidth,-0.15 * \dSwidth);
   \coordinate(rc) at (.32*\dSwidth,-0.15 * \dSwidth);
   \coordinate(k1) at (-.35*\dSwidth,0);
   \coordinate(k2) at (.35*\dSwidth,0);
   \draw [dashed](k1) .. controls (lc) ..  (L);
   \draw (R) arc [start angle=0, end angle=180, radius=.2*\dSwidth];
   \draw (R) arc [start angle=0, end angle=-180, radius=.2*\dSwidth];
   \draw [dashed](R) .. controls (rc) .. (k2);
\end{dSdiagram}
\quad &< \quad
\begin{dSdiagram}[aspect=0.15, width=2.8cm]
   \coordinate(L) at (-0.2*\dSwidth,-0.45 * \dSwidth) ;
   \coordinate(R) at (0.2*\dSwidth, -0.45 * \dSwidth) ;
   \coordinate(k1) at (-.35*\dSwidth,0);
   \coordinate(k2) at (.35*\dSwidth,0);
   \draw [dashed](k1) .. controls (L) and (R) .. (k2);
\end{dSdiagram}
    \end{align}
    is much more stringent. Due to the enhanced long-short coupling, the inflaton loop contributes non-trivially to the tachyon propagator, with two copies of the enhancement factor, leading to
    \begin{align}
        \frac{1}{(4\pi)^2}(\lambda H)^2\left(\frac{k_\mu}{k_{\text{CMB}}}\right)^{2\alpha}<1~.\label{chiPropPertConstraint}
    \end{align}
    With a large scale hierarchy $k_\mu/k_{\text{CMB}}\simeq 10^7$, the triple coupling is tightly constrained for a non-zero choice of $\alpha$.

    \item \textit{Null detection of the $(T,E)^3$-bispectrum at $\alpha\to 0$}. As mentioned before, the $\alpha\to 0$ limit of our signal is degenerate with local non-Gaussianity, which has been constrained by \Planck\ 2018 to be $f_{\rm NL}^{\rm loc}<5.1~(68\% \text{CL})$ \cite{Planck:2019kim} for temperature and $E$-mode polarisation maps. Using $\mathcal{C}(3/2)=1/4$, we thus obtain a constraint on the size of the product of triple and mixing couplings,
    \begin{align}
        (\lambda H)(\rho H)<\frac{6}{5}\times 5.1\times 2\pi \Delta_\zeta\simeq 0.002~.
    \end{align}

    \item \textit{Null detection of the $(T,E)^4$-trispectrum at $\alpha\to 0$}. In addition to the bispectrum, the triple coupling also induces a four-point function that is degenerate with the local-shape template (parametrised by the amplitude estimator $\tau_{\rm NL}^{\rm loc}$) when $\alpha\to 0$. The null detection of $\tau_{\rm NL}^{\rm loc}$ therefore also poses a constraint on the size of the triple coupling. Following the same computations as in the previous subsection, we arrive at
    \begin{align}
        \nonumber B_4^\zeta &=
 \begin{dSdiagram}[aspect=0.15, width=3.5cm]
   \node(L) at (-0.25*\dSwidth,-0.35 * \dSwidth) [vertex];
   \node(R) at (0.25*\dSwidth, -0.35 * \dSwidth) [vertex];
   \coordinate(k1) at (-.35*\dSwidth,0);
   \coordinate(k2) at (-.15*\dSwidth,0);
   \coordinate(k3) at (.15*\dSwidth,0);
   \coordinate(k4) at (.35*\dSwidth,0);
   \draw (k1) --  (L) node[midway, left]{$k_1$};
   \draw (k2) --  (L) node[midway, right]{$k_2$};
   \draw [dashed] (L) -- (R) node[midway, above]{$s$};
   \draw (R) -- (k3) node[midway, left]{$k_3$};
   \draw (R) -- (k4) node[midway, right]{$k_4$};
\end{dSdiagram} \ +\ \text{2 perms}\\
        &\approx(2\pi^2\Delta_\zeta^2)^3\, \tau_{\rm NL}^{(\alpha)} \,\left[\frac{4}{k_1 k_2 k_3 k_4 k_{12} k_{34} s^3}\left(\frac{k_{12}}{2s}\right)^{\alpha}\left(\frac{k_{34}}{2s}\right)^{\alpha}+\text{2 perms}\right]~,
    \end{align}
    with
    \begin{align}
        \tau_{\rm NL}^{(\alpha)} =-\frac{2^{\tilde{\nu}-5/2}}{\pi^2 \Delta_\zeta^2} H^2\lambda^2\, \mathcal{C}(\alpha+3/2)~.\label{tauNLDef}
    \end{align}
    The $\alpha\to 0$ limit matches the local template for
    \begin{align}
        \tau_{\rm NL}^{(\alpha)}=4\tau_{\rm NL}^{\rm loc}~.
    \end{align}
    Using the \Planck\ 2013 constraint $\tau_{\rm NL}^{\rm loc}<2800~(95\%\text{CL})$ \cite{Planck:2013wtn}, we arrive at an observational upper bound for the triple coupling,
    \begin{align}
        \lambda H< \sqrt{8\times 4\times 2800}\,\pi \Delta_\zeta\simeq 0.04~.
    \end{align}
    Note that this is complementary to the theoretical constraint from loop perturbativity \eqref{chiPropPertConstraint} for different choices of $\alpha$.

    \item \textit{Control of non-Gaussianity}. The growth of the cosmological tachyon collider signal in the bispectrum and trispectrum suggests increasing deviation from Gaussian statistics. In order to ensure the validity of Gaussian statistics as a starting point for the perturbative expansion, we need to impose constraints on the size of eventual non-Gaussianities. In the case of local non-Gaussianity ($\alpha=0$), the corresponding constraints are typically \cite{Celoria:2021vjw}
    \begin{align}
        |f_{\rm NL}^{(0)}|\, \Delta_\zeta&<1~,\\
        |\tau_{\rm NL}^{(0)}|\, \Delta_\zeta^2&<1~,
    \end{align}
    where $\Delta_\zeta$ was defined in Footnote \ref{foot}. In cases with long-short enhancement, these bounds effectively translate to
    \begin{align}\label{pertbound1}
        |f_{\rm NL}^{(\alpha)}|\, \Delta_\zeta \left(\frac{k_\mu}{k_{\text{CMB}}}\right)^{\alpha}&<1~,\\
        |\tau_{\rm NL}^{(\alpha)}|\, \Delta_\zeta^2 \left(\frac{k_\mu}{k_{\text{CMB}}}\right)^{2\alpha}&<1~.\label{pertbound2}
    \end{align}
    Using the definitions of the non-Gaussianity estimators, i.e. \eqref{fNLDef} and \eqref{tauNLDef}, we obtain
    \begin{align}
       \frac{2}{\pi}(\lambda H)(\rho H)\,\mathcal{C}(\alpha+3/2)\left(\frac{k_\mu}{k_{\text{CMB}}}\right)^{\alpha}&<1~, \label{ControlNonG3pt}\\
        \frac{2^{\tilde{\nu}-5/2}}{\pi^2} (\lambda H)^2\, \mathcal{C}(\alpha+3/2) \left(\frac{k_\mu}{k_{\text{CMB}}}\right)^{2\alpha}&<1~.\label{ControlNonG4pt}
    \end{align}
    Notice that with $0<\alpha<1$, these constraints are roughly degenerate with the perturbativity bounds \eqref{MixingPertConstraint} and \eqref{chiPropPertConstraint}, as required by the consistency of perturbation theory.
    
\end{itemize}


\section{Non-Gaussian templates}\label{sec:TemplateSection}

In order to compare the constraints on our signal from CMB $(T,E)^3$ and $\mu T$ correlations, it is necessary to consider the shape of the bispectrum not just in the squeezed limit but for arbitrary kinematics.
The analytic approach of the previous section provides the bispectrum in the squeezed limit alone, which suffices to predict the $\mu T$ correlation only.
To obtain the full shape, one could compute the tree-level bispectrum numerically, either evaluating the leading Feynman diagram by numerical quadrature or using a computational package designed for calculating correlators, such as \texttt{CosmoFlow} \cite{Pinol:2023oux,Werth:2023pfl}. 
Here, we follow a simpler approach: since we expect the bispectrum to be strongly peaked in the squeezed limit, most of the constraining power of CMB $(T,E)^3$ observations comes from squeezed triangles.
Therefore, any shape function that is symmetric in the external kinematics, has the correct squeezed limit, and has no other peaks should approximate our full bispectrum well, at least for the purpose of a signal-to-noise estimate.

Here, we present three possible shape functions.
In Section \ref{sec:PixiePlanck SN}, we show that, as expected, they have very similar signal-to-noise ratios in CMB data, even though their amplitudes in the equilateral limit are somewhat different.
The first, Shape 1, uses the following squeezed-limit asymptotics\footnote{The variables $k_T \equiv k_1 + k_2 + k_3$, $e_2 \equiv k_1 k_2 + k_2 k_3 + k_3 k_1$, and $e_3 \equiv k_1 k_2 k_3$ are elementary symmetric polynomials in the external momenta.}

\begin{align}
   \frac{4 e_3}{k_T^2} &\sim k_L\,, & \frac{k_T}{2} &\sim k_S,
\end{align}
where $k_L$ is the smallest side of the squeezed triangle, and $k_S \gg k_L$ is size of the two other sides. Substituting these symmetric expressions in the squeezed limit bispectrum we obtain a full shape with the correct squeezed limit:
\begin{equation}
   B_3^{(S1)} \equiv  2^{-3(1 + \alpha)} (2\pi^2 \Delta_\zeta^2)^2 e_3^{-2} \cdot \fnlA \left( \frac{k_T^3}{e_3} \right)^{1 + \alpha}.
   \label{eq:Shape 1}
\end{equation}
Alternatively, Shape 2 uses the following squeezed-limit asymptotics
\begin{align}
   \frac{e_3}{e_2} &\sim k_L \,, &\sqrt{e_2} &\sim k_S\,,
\end{align}
to write the bispectrum as
\begin{equation}
   B_3^{(S2)} \equiv  (2\pi^2 \Delta_\zeta^2)^2 e_3^{-2} \cdot \fnlA \left( \frac{e_2^{3/2}}{e_3} \right)^{1 + \alpha}.
   \label{eq:Shape 2}
\end{equation}
Finally, Shape 3 consists of a sum over permutations of external kinematics in equation \eqref{B3zetaApprox}, so that a bispectrum template with the correct squeezed limit is found to be
\begin{equation}
   B_3^{(S3)} \equiv 2^{1 - \alpha} \cdot (2\pi^2 \Delta_\zeta^2)^2 e_3^{-2} \cdot \fnlA \left( \frac{k_1 k_2}{k_{12}^2} \left( \frac{k_{12}}{k_{3}} \right)^{\alpha + 1} + 2\,\text{perms} \right).
   \label{eq:Shape 3}
\end{equation}
In other words, this shape is reached by approximating the full bispectrum by its squeezed limit \eqref{B3zetaApprox}.

\section{Prospects for the $\mu T$ correlation}\label{muTProspectsSection}

The bispectrum \eqref{SqueezedLimitBispectrum} grows faster than local-type non-Gaussianity in the squeezed limit.
Correlations between CMB $\mu$ distortions and temperature anisotropies are sensitive to the squeezed limit of the bispectrum, so the amplitudes of bispectra with this squeezed behaviour are constrained by $\mu T$ correlations.
The relation between primordial non-Gaussianity and $\mu T$ correlations was first described in \cite{Pajer:2012vz}, and although more refined and accurate treatments have since been developed, see for example \cite{Pajer:2012qep,Ganc:2012ae,Biagetti:2013sr,Chluba:2016aln,Remazeilles:2018kqd,Chluba:2022xsd,Chluba:2022efq,Kite:2022eye}, we follow closely the approach of \cite{Pajer:2012vz} to estimate the $\mu T$ correlation and forecast the constraining power of $\mu T$ measurements on the amplitude of the squeezed bispectrum.
First, let's very briefly review the physics of $\mu$ distortions.

\subsection{CMB spectral distortions}
In the electron-photon-baryon plasma that fills the universe following the hot Big Bang, thermal equilibrium is maintained by scattering processes between charged particles and the electromagnetic field. Double Compton scattering, single Compton scattering, and Bremsstrahlung are the dominant processes.
The efficiency of these processes falls with the temperature of the plasma, and between redshifts of approximately $z \approx 2 \times 10^6$ and $z \approx 5 \times 10^4$ (the $\mu$ era), only single Compton scattering is non-negligible except at very long wavelengths. Photon number is conserved in single Compton scattering, so the radiation filling the universe is approximated by a Bose-Einstein distribution with a chemical potential $\mu$. When the chemical potential is small compared to the temperature, the change $\delta \mu$ due to a fractional change $\delta E/E$ in the energy in the radiation is given by
\begin{equation}
   \delta \mu \approx 1.4 \frac{\delta E}{E}.
\end{equation}
Energy is dissipated into the radiation by diffusion damping of acoustic waves in the plasma. It has been recognized for a very long time that the $\mu$ monopole is a powerful probe of the short-scale primordial power spectrum \cite{Sunyaev:1970plh,Hu:1994bz}, and this provides one of the strongest reasons to improve our measurement of the CMB spectrum \cite{Cabass:2016giw,Chluba:2016bvg}. 
The amplitude of acoustic waves features a damping factor $e^{-k^2/k_D^2(t)}$, where the time-dependent damping scale $k_D$ is approximated by \hbox{$k_{D,\, \mathrm{i}} \approx 1.1\times 10^4 \, \mathrm{Mpc}^{-1}$}  at the beginning of the $\mu$ era, and by $k_{D,\, \mathrm{f}} \approx 46 \, \mathrm{Mpc}^{-1}$ at its end.  The energy density in the acoustic modes is proportional to their squared amplitude, so (see \cite{Pajer:2012vz} for details) the energy dissipated by diffusion damping determines the multipole moments of the $\mu$ distortion through
\begin{equation}
   a^{\mu}_{lm} \approx 9.2 \pi (-1)^l \int \frac{\dd^3 k_1 \dd^3 k_2}{(2\pi)^6} Y^*_{lm} ({\hat k}_+) \zeta(\vb k_1) \zeta(\vb k_2) W\left( \frac{k_{+}}{k_\text{scatter}} \right) j_l(k_{+} \, r_\text{LS}) \left[ e^{-\left( k_1^2 + k_2^2 \right)/ k_D^2} \right]_\mrf^\mri, \label{eq:a mu}
\end{equation}
where $j_l$ is a spherical Bessel function, $r_\text{LS}$ is the comoving distance to the surface of last scattering, $\zeta$ is the comoving curvature perturbation, and $W$ is the Fourier transform of a top-hat function of size $k_\text{scatter}$ which models redistribution of energy by scattering.
The vector ${\vb k}_+ = \vb k_1 + \vb k_2$.
On large scales, the temperature anisotropy $T$ is approximately related to the comoving curvature perturbation by
\begin{equation}
   a^T_{lm} \approx \frac{4 \pi}{5} (-1)^l \int \frac{\dd^3 k_3}{(2\pi)^3}  Y^*_{lm} ({\hat k}_{3}) \zeta(\vb k_3) j_l(k_3 \, r_{\text{LS}}),
\end{equation}
so that the $\mu T$ correlation takes the following form
\begin{align}\nonumber
   C^{\mu T}_l &= \ev{(a^T_{lm})^* a^\mu_{lm})} \\
&\approx 9.2 \pi \frac{4 \pi}{5}  \int \frac{\dd^3 k_1 \dd^3 k_2}{(2\pi)^6} |Y_{lm} ({\hat k}_{+})|^2 W\left( \frac{k_{+}}{k_\text{scatter}} \right)  \left[ e^{-\left( k_1^2 + k_2^2 \right)/ k_D^2} \right]_\mrf^\mri \nonumber \\  
& \quad    j_l(k_{+} \, r_\text{LS})^2  B_3^\zeta(k_1, k_2, k_{+}) \, .
\end{align}
Changing variables to $\vb k_+ = \vb k_1 + \vb k_2$ and $\vb k_- = \vb k_1 - \vb k_2$, the integral becomes
\begin{align}
   C^{\mu T}_l &\approx \frac{9.2 \pi^2 }{10}  \int \frac{\dd^3 k_+ \dd^3 k_-}{(2\pi)^6}  |Y_{lm} ({\hat k}_{+})|^2 W\left( \frac{k_{+}}{k_\text{scatter}} \right) \left[ e^{-\left( k_+^2 + k_-^2 \right)/ 2 k_D^2} \right]_\mathrm{f}^\mathrm{i}  \nonumber\\
   & \quad j_l(k_{+} \, r_\text{LS})^2 \,  B_3^\zeta\left( \frac{k_-}{2}, \frac{k_-}{2}, k_{+}\right)  \, ,
   \label{eq:mu T general}
\end{align}
where the approximation $\norm{\vb k_+ \pm \vb k_-} \approx k_-$ has been used.
This is a result of the spherical Bessel function peaking around CMB scales and the difference of damping Gaussians $\left[ e^{-\left( k_+^2 + k_-^2 \right)/ 2 k_D^2} \right]_\mathrm{f}^\mathrm{i}$ peaking on much shorter scales.


\subsection{Spectral distortions from the cosmological tachyon collider}
The squeezed bispectrum \eqref{SqueezedLimitBispectrum} has the following approximate form:
\begin{equation}
   B_3^\zeta\left(k_S, k_S, k_L \right) \approx (2\pi^2 \Delta_\zeta^2)^2 \, \fnlA \frac{1}{(k_L k_S)^3} \left( \frac{k_S}{k_L} \right)^\alpha,
\end{equation}
where $k_L \ll k_S$, $\alpha = \tilde \nu - 3/2$ is determined by the growth rate of the transient tachyon.
The integrals appearing in equation \eqref{eq:mu T general} for $C_l^{\mu T}$ therefore factorise.
The integral over long modes $k_+$ is simplified by noting that the spherical Bessel functions decay at scales much shorter than $l/r_\text{LS}$, and $k_D$ and $k_\text{scatter}$ are much shorter than this scale.
This integral therefore takes the form of a Weber-Schafheitlin discontinuous integral \cite{NIST:DLMF}
\begin{align}
   \int k_+^2 \dd k_+  j_l(k_{+} \, r_\text{LS})^2 k_+^{-3 - \alpha}  e^{-k_+^2/2k_D^2} W\left( \frac{k_{+}}{k_\text{scatter}} \right)  &\approx \int k_+^2 \dd k_+  j_l(k_{+} \, r_\text{LS})^2 k_+^{-3 - \alpha} \\
   &= \frac{\pi}{4 r_{\text{LS}}} \frac{\left( \frac{r_\text{LS}}{2} \right)^{1 + \alpha} \Gamma\left( l  - \alpha/2 \right) \Gamma(2 + \alpha)}{ \Gamma\left( \frac{3 + \alpha}{2} \right)^2 \Gamma\left( l + 2 + \alpha/2 \right) }\,, \label{eq:WSDI}
\end{align}
subject to the condition $-2 < \alpha < l$ for convergence.
Since $3/2 < \tilde \nu < 5/2$, this condition is always satisfied.
The integral over short modes $k_-$ consists of powers of $k_-$ multiplying a Gaussian; the result is a gamma function
\begin{equation}
   \int  k_-^2 \dd k_- 2^{3 - \alpha} k_-^{-3 + \alpha} e^{-k_-^2/2k_D^2} = 2^{2 - \alpha} \left( \sqrt{2} \ k_D \right)^{\alpha} \Gamma\left( \frac{\alpha}{2} \right), \label{eq:mu gamma}
\end{equation}
so that the $\mu T$ correlator becomes
\begin{equation}
   C^{\mu T}_l \approx (2\pi^2 \Delta_\zeta^2)^2 \fnlA \cdot 2^{-4-\alpha}  \frac{9.2 }{5 \pi^2 \alpha}  \left[\left( \frac{k_D \, r_\text{LS}}{\sqrt{2}} \right)^\alpha \right]_\mrf^\mri \frac{ \Gamma(2 + \alpha) \Gamma\left( l  - \alpha/2 \right)}{\Gamma \left(\frac{3 + \alpha}{2} \right) \Gamma\left( l + 2 + \alpha/2 \right)}.
   \label{eq:mu T correlator}
\end{equation}
Note that there is a removable discontinuity at $\alpha=0$ due to a divergence in the integral over the short modes which cancels when the difference between the integrals at the beginning and end of the $\mu$ era is taken. Moreover, we should comment on the $l$-dependence of this result. Stirling's asymptotic formula for the gamma function implies that these correlators scale as $l^{-2 - \alpha}$ for large $l$.
As $0 < \alpha < 1$, the correlators decay no faster than $l^{-3}$.
This is in contrast to what is expected for a scale-invariant two-point correlation, which scales as $1/l(l+1)$. This highlights that while the primordial curvature perturbations are approximately scale-invariant, the resulting $\mu$ distortions are not.
One way to understand this is that $\mu$ distortions are imprinted during a period between inflation and recombination which is determined by the physics of photon scattering in the electron-proton plasma.
Therefore, the statistics of the $\mu$ distortions may depend on the length- and time-scales intrinsic to these processes as well as external momenta --- the result found in \cite{Pajer:2012vz} in the case of local-type non-Gaussianity represents an exception in which scale invariance is restored.

Using equations \eqref{eq:a mu}, \eqref{eq:WSDI}, and \eqref{eq:mu gamma}, it is not too difficult to calculate the non-Gaussian contribution to the $\mu$ angular power spectrum
\begin{equation}
   C^{\mu\mu}_l = (2\pi^2 \Delta_\zeta^2)^3 \tau_\mathrm{NL}^{(\alpha)} \cdot 2^{-4 - 2\alpha} \frac{(9.2\pi)^2}{(2\pi)^6} \left[\eval{ \left(\frac{k_D \, r_\text{LS}}{\sqrt{2}}\right)^\alpha }_\mrf^\mri \, \right]^2 \frac{\Gamma\left( \frac{\alpha}{2} \right)^2 \Gamma(l - \alpha)\, \Gamma(2 + 2\alpha)}{ \Gamma\left( \frac{3}{2} + \alpha \right)^2 \Gamma(l + 2 + \alpha)} \, .
   \label{eq:C mu mu NG}
\end{equation}
Taking $\alpha=0.8$, $l=2$, $\fnlA \sim 10^{-2}$, and $\tau_\mathrm{NL}^{(\alpha)} \sim 10^{-3} $ as a representative region of parameter space satisfying the constraints of Section \ref{sec:parameter constraints} and maximising the observable signals, $C_{2}^{\mu T} \sim 10^{-14}$, and $C_2^{\mu \mu} \sim 10^{-18}$.
Although the non-Gaussian contribution to the $\mu$ angular power spectrum is as usual large compared to its Gaussian part, it is nevertheless small compared to instrumental noises in measurements of $\mu$ for the experiments considered in the following section. Although a direct detection of the $\mu$ monopole or power spectrum produced by our class of models remains out of reach for current experiments, in the absence of atmospheric noise, the CMB-S4 experiment would reach the requisite sensitivity \cite{CMB-S4:2023zem}. This raises the possibility that future ground- or space-based observatories may constrain our models through $\mu$ and $\mu^2$ measurements.

\begin{figure}[h!]
   \centering
   \includegraphics[width=0.8\textwidth]{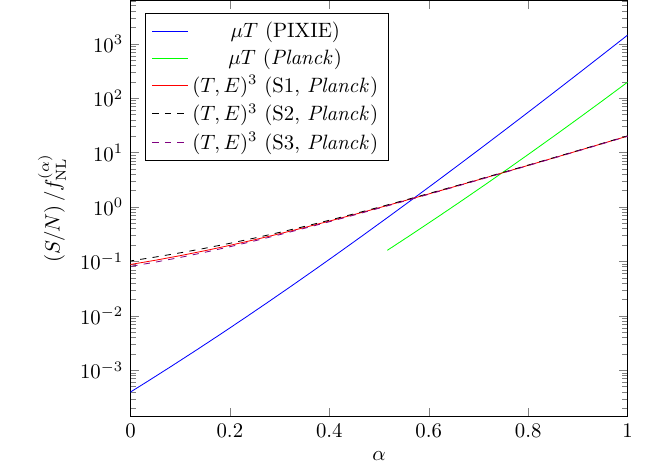}
   \caption{Signal-to-noise for $\fnlA$ as function of $\alpha$, both defined in \eqref{SqueezedLimitBispectrum}. The \Planck\ $\mu T$ line is shown only for $\alpha > 0.5$ because for smaller $\alpha$, much of the signal comes from $l > 100$, where the use of the Sachs-Wolfe transfer function to estimate the signal and noise becomes a poor approximation. The signal-to-noise in both the CMB and $\mu T$ correlations is proportional to $\fnlA$.  The three shape functions used in the \Planck\ $(T,E)^3$ analysis are defined in equations \eqref{eq:Shape 1}, \eqref{eq:Shape 2}, and \eqref{eq:Shape 3} respectively, and give very similar results, as expected.} 
   \label{fig:signal-to-noise}
\end{figure}


\subsection{Signal-to-noise with \Planck, PIXIE and CMB-S4}
\label{sec:PixiePlanck SN}

In this subsection we discuss our estimates of the signal-to-noise ratio for the amplitude $\fnlA$ of non-Gaussianity from CMB temperature and polarization anisotropies and for the cross correlation of $\mu$ and $T$ anisotropies, where we use \Planck, PIXIE, and CMB-S4 as benchmark experiments. 

\paragraph{PIXIE}
Following \cite{Pajer:2012vz}, we use a Fisher forecast to place a lower bound on the variance of a measurement  of the amplitude $\fnlA$ of the squeezed bispectrum through $\mu T$ correlations.
This implies that, if the multipole moments of the $\mu T$ correlation have an approximately Gaussian likelihood, the signal-to-noise ($S/N$) in a measurement of $\fnlA$ is at most
\begin{equation}
   (S/N)^2 \lesssim \sum_l \frac{(2l + 1) \left(C^{\mu T}_l\right)^2}{C^{\mu\mu, \, \mathrm N}_l \, C^{T T}_l}\,,
   \label{eq:SN general}
\end{equation}
for an experiment limited by cosmic variance in measurements of $T$ and by noise $ C^{\mu\mu, \, \mathrm N}_l$ in measurements of $\mu$ (this includes instrumental and foreground effects).
As described below equation \eqref{eq:C mu mu NG}, for our signal, the $\mu$ angular power spectrum is typically of order $10^{-18}$ or smaller, so instrumental noise dominates over cosmic variance in the $\mu$ measurements.
For a PIXIE-like experiment, the noise takes the approximate form 
\begin{equation}
   C^{\mu\mu, \, \mathrm N}_l \approx w_\mu^{-1} \, e^{l^2/l_\text{max}^2},
\end{equation}
with $l_{\text{max}} \approx 84$ determined by the beam width and $w_{\mu}^{-1} \approx 4\pi \times 10^{-16}$ the sensitivity, and
\begin{equation}
   C_l^{T T} \approx \frac{2\pi \Delta_\zeta^2 }{25l(l+1)} \, .
\end{equation}
Therefore, using the expression \eqref{eq:mu T correlator} for $C_l^{\mu T}$,
\begin{samepage}
\begin{align}
   (S/N)^2 & \approx \left[ 2^{-4-\alpha}  \frac{9.2 (2\pi^2 \Delta_\zeta^2)^2 \fnlA}{5 \pi^2 \alpha}  \left[\left( \frac{k_D \, r_\text{LS}}{\sqrt{2}} \right)^\alpha \right]_\mrf^\mri \frac{\Gamma(2+\alpha)}{\Gamma(\frac{3 + \alpha}{2})}  \right]^2 \cdot \frac{25}{2\pi \Delta_\zeta^2 w_{\mu}^{-1}}   \nonumber  \\
   & \quad \sum_l l(l+1)(2l+1) \left( \frac{\Gamma(l - \frac{\alpha}{2})}{\Gamma(l + 2 + \frac{\alpha}{2})} \right)^2 e^{-(l / 84)^2} \, .
   \label{eq:mu T S/N as a sum}
\end{align}
\end{samepage}
When numerical values of this signal-to-noise are required below, they are computed by numerically summing from $l=2$ to $l=100$.
Generally, very little of the signal comes from $l$s as large as 100 due to the $l^{-2-\alpha}$ asymptotic scaling of $C_l^{\mu T}$ and the rapid growth of the noise $C_l^{\mu \mu,\,N}$.
In fact, when $\alpha$ is not close to zero and the noise in $\mu$ is not decreasing, the first few $l$s dominate the signal-to-noise.
Therefore, as long as sensitivity is maintained for $l \lesssim 10$, the sensitivity of an experiment is much more important than its spatial resolution when constraining our signal.

\paragraph{\Planck} In order to determine whether our signal is better-constrained by $\mu T$ correlations than by CMB anisotropies alone, it is necessary to estimate the signal-to-noise of our bispectrum in \Planck\ data.
We made this estimate for each of the template bispectra in Section \ref{sec:TemplateSection}, verifying that the difference in signal-to-noise among the templates is negligible. 
The $S/N$ was estimated using Fisher information computed using the \texttt{cmbbest} Python package \cite{Sohn:2023fte}.
Given shape functions for the Bardeen potential $\Phi$, which is approximated on super-Hubble scales by
\begin{equation}
   \Phi \approx \frac{3}{5} \zeta \, ,
\end{equation}
\texttt{cmbbest} estimates the amplitudes of the shapes in the primordial bispectrum  and the Fisher information matrix for those amplitudes using \Planck\ $T$ and $E$ data.

\begin{figure}[t]
   \centering
   \includegraphics{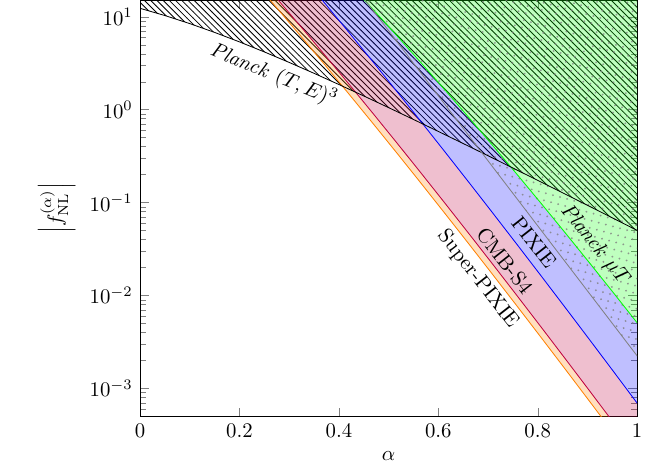}
   \caption{Minimum detectable $\fnlA$ for PIXIE and \Planck\ observations, and a CMB-S4-like experiment with a conservative noise model. The striped area indicates the region of parameter space already excluded by \Planck\ $(T,E)^3$ observations. A point $(\alpha,\fnlA)$ is marked as detectable if it provides a signal-to-noise ratio greater than unity. The plot additionally includes a proposed `Super-PIXIE' experiment \cite{Chluba:2019nxa,Kogut:2019vqh}, which would deliver a sensitivity close to an order of magnitude better than PIXIE's. The grey dotted region corresponds to the perturbativity constraint \eqref{pertbound1}. In this region, the primordial curvature perturbation becomes substantially non-Gaussian on $\mu$ scales, and our perturbative calculations become unreliable.}
   \label{fig:fnl constraint}
\end{figure}

\paragraph{\Planck\ $\mu T$} In \cite{Rotti:2022lvy}, an interesting analysis of $\mu T$ and $\mu E$ correlations was performed using a $\mu$ map reconstructed from \Planck\ data (see \cite{Khatri:2015tla} for an earlier analysis).
Foregrounds and instrumental noises were treated carefully in order to constrain the amplitude $\fnl^{\rm loc}$ of local non-Gaussianity, finding $|\fnl^{\rm loc}| \lesssim 6800$ at a 95\% confidence level.
To make a crude Fisher estimate of the constraining power of $\mu T$ correlations from \Planck\ for the signal \eqref{SqueezedLimitBispectrum}, we note that the noise in the reconstructed \Planck\ $\mu$ map in \cite{Rotti:2022lvy} was dominated by foregrounds at $l \lesssim 200$.
The noise due to foregrounds took the approximate form
\begin{equation}
   C^{\mu\mu,\,N}_l \approx  5 \times 10^{-13} \cdot l^{-1.38} \, .
\end{equation}
Because the $\mu$ noise has a minimum around $l \approx 250$, it turns out that for $\alpha \lesssim 0.5$, much of the signal-to-noise in the Fisher forecast comes from $l > 100$.
In the present work, calculations of the $\mu T$ signal and $TT$ noise use approximate transfer functions which are less accurate for these $l$s.
Therefore, in the plots that follow, the estimated signal-to-noise in $\fnlA$ from the \Planck\ $\mu T$ correlation is not shown at low $\alpha$.

\paragraph{CMB-S4} The possibility of detecting local-type non-Gaussianity through anisotropic $\mu T$ correlations with the upcoming CMB-S4 experiment was considered in \cite{CMB-S4:2023zem}.
The constraining power depends on the degree of correlation between frequency bands in the atmospheric noise, with even a small uncorrelated component substantially increasing the noise in the $\mu$ anisotropies.
Using a conservative noise model based on \cite{CMB-S4:2023zem}, assuming constant noise $C^{\mu\mu,\,N}_l = 10^{-16}$, we consider the ability of a CMB-S4-like experiment to constrain our signal.
This constant noise model is more pessimistic even than the case where atmospheric noise is uncorrelated between frequency bands; at small $\alpha$, this represents a conservative forecast, whereas as $\alpha$ approaches 1, this provides a good approximation of the  signal-to-noise.

Figure \ref{fig:signal-to-noise} shows the Fisher estimated signal-to-noise in the parameter $\fnlA$ as a function of $\alpha$ for some of the instrument and observable combinations discussed above --- $\mu T$ correlations with PIXIE, CMB $(T,E)^3$ correlations with \Planck, and $\mu T$ correlations with \Planck.
Figure \ref{fig:fnl constraint} shows the regions of the $(\alpha, \fnlA)$ parameter space that would produce at least a marginally detectable signal (signal-to-noise ratio greater than unity) in each of these instrument-observable combinations described here.
Alongside the experiments described above, this figure also includes regions corresponding to  a proposed `Super-PIXIE' experiment \cite{Chluba:2019nxa,Kogut:2019vqh}, which would use multiple spectrometers in order to constrain foregrounds better and provide sensitivity around an order of magnitude higher than PIXIE's.


\section{Conclusions and outlook} \label{ConclusionsSect}
\begin{figure}[t!]
   \centering
   \includegraphics[width=0.9\textwidth]{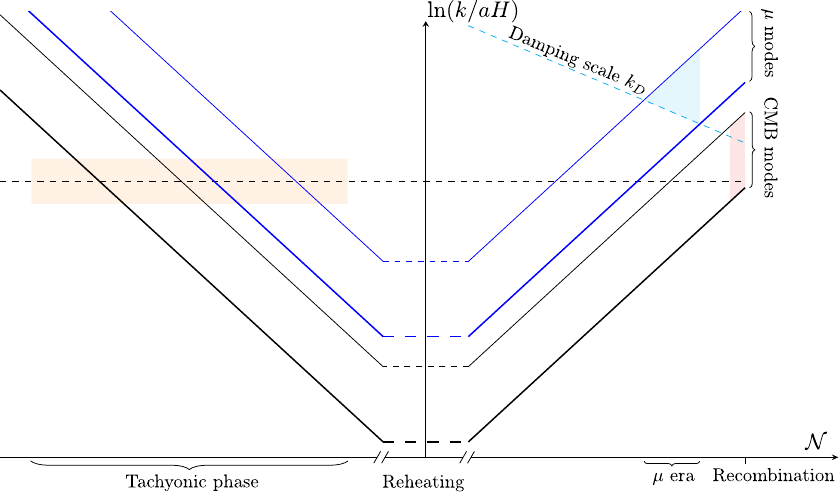}
\caption{A pictorial summary of the cosmological tachyon collider. This plot shows the evolution history of different scales and how the tachyonic enhancement in the \textit{early} universe (left half) becomes encoded in cosmological observables in the \textit{late} universe (right half). The horizontal axis represents the flow of time in efolds and the vertical axis gives the logarithm of the ratio of physical momenta and the Hubble scale at relevant times. The \textcolor{black}{black} dashed line at $\ln(k/aH)=0$ represents horizon crossing. The \textcolor{black}{black} and \textcolor{blue}{blue} lines correspond the largest (thick) and smallest (thin) scales of modes of temperature anisotropies and $\mu$ distortions, respectively. The \textcolor{cyan}{cyan} dashed line gives the diffusion damping scale above which fluctuations are dissipated. The \textcolor{orange}{orange} region indicates the interaction with the transient tachyon during inflation. The resultant signal can be probed by $(T,E)^3$ correlators living in the \textcolor{pink}{pink} region or the $\mu T$ correlators across the \textcolor{cyan}{cyan} and \textcolor{pink}{pink} regions, which enjoy a larger momentum hierarchy and therefore a stronger signal.}
   \label{ModeHistoryPlot}
\end{figure}

We conclude with an illustration of our physical picture in Figure \ref{ModeHistoryPlot}. In this work, we have built and studied a series of models that generate a primordial non-Gaussian signal from inflation that strongly couples short and long-scale perturbations. The observable of interest is the squeezed limit of the bispectrum of curvature perturbations. We have shown that in this limit, a signal is found of the form
\begin{align}
	\ex{\zeta(k_S)\zeta(k_S)\zeta(k_L)}'=(2\pi^2\Delta_\zeta^2)^2 \,f_{\rm NL}^{(\alpha)}\, \frac{1}{k_S^3 k_L^3}\left(\frac{k_S}{k_L}\right)^\alpha~,\quad \alpha>0~\,.
\end{align}
where $\alpha=0$ corresponds to the well-known local non-Gaussianity. For $\alpha >0$ one finds that the signal peaks in the most separated scales that one can observationally access. We have shown that this signal can be large and is currently bounded, in different regions of parameter space, either by CMB anisotropies or by the cross correlation of CMB temperature anisotropies with spectral distortion anisotropies of the $\mu$ type. Our study shows that there are classes of consistent models of inflation for which spectral distortions will provide the leading observational bounds already with planned experiments! For the models we considered here, the constraining power of spectral distortion anisotropies will become even more dominant over temperature anisotropies in the future. On the theoretical side, our study provided a smoking gun signal for the presence of (temporarily) tachyonic spectator fields during inflation. Our work is the natural extension of cosmological collider physics to negative masses, which are not usually considered due to their dangerous instabilities. We have shown that such instabilities can be kept under control while still obtaining an observationally large signal.

Our work opens up several avenues for future research:
\begin{itemize}
   \item It would be nice to perform a much more careful forecast of the constraints that current and future experiments can put on our class of models, including more realistic assumptions about noises, foregrounds and the inhomogeneous thermal history, perhaps along the lines of \cite{Remazeilles:2018kqd,Rotti:2022lvy,CMB-S4:2023zem}. 
    \item The ultra-squeezed primordial non-Gaussianity produced by our models would also have observable consequences in galaxy surveys, including scale-dependent bias alongside a contribution to the bispectrum of galaxies. It is possible that these large-scale structure observables as well as others such as 21-cm could provide constraints on our class of models.
    \item A scaling in the squeezed limit that is faster than that of local non-Gaussianity signals an instability. Moreover, an ever increasing coupling between long and short modes is a serious threat to the principle of separation of scales, the resting pillar of effective field theories. It would be nice to obtain general, model independent bounds on the growth in this limit, perhaps employing ideas from quantum information.
    \item Several modifications and improvements of our setup are possible. To name a few, one can consider testing the long-short coupling in the tensor sector \cite{Orlando:2021nkv,Ota:2014hha}, constructing a UV completion for the mass filter in Model II, and going beyond the Gaussianity constraints \eqref{ControlNonG3pt} and \eqref{ControlNonG4pt} using non-perturbative techniques \cite{Celoria:2021vjw,Creminelli:2024cge}.
    \item Given that the cosmological tachyon collider signal peaks towards the squeezed limit, its detectability crucially relies on the measurement precision of low-$l$ modes despite their cosmic variance (which is relevant only if they are actually detected). The very existence of such faster-than-local shape non-Gaussianity calls for the design of future $\mu$-distortion experiments that excel at low-$l$ sensitivity.
\end{itemize}
We hope that our work will contribute to building an even more compelling science case for a future mission targeting the spectrum of the cosmic microwave background.


\paragraph{Acknowledgements} We would like to thank Xingang Chen and Jens Chluba for useful discussions. We are also grateful to the organizers and participants of the UK Spectral Distortions Meeting 2023, an event that prompted our investigation. E.P. has been supported in part by the research program VIDI with Project No. 680-47-535, which is (partly) financed by the Netherlands Organisation for Scientific Research (NWO). C.M.~is supported by Science and Technology Facilities Council (STFC) training grant ST/W507350/1. This work has been partially supported by STFC consolidated grant ST/T000694/1 and ST/X000664/1 and by the EPSRC New Horizon grant EP/V017268/1. 

\newpage
\appendix

\bibliographystyle{JHEP}
\bibliography{CTCRefs}

\end{document}